\documentclass[12pt,tightenlines,eqsecnum,floats,aps,amsmath,amssymb,nofootinbib,prd,shownopacs]{revtex4}

\usepackage{amsmath,amsthm,latexsym,amssymb,amsfonts}
\usepackage{enumerate}
\usepackage{graphicx}
\usepackage{calc}
\usepackage[mathscr]{eucal}

\usepackage{LQC-symbols}

\newcommand{\Fou}{\mathcal{F}} 
\newcommand{\defi}{\mathcal{K}} 
\newcommand{\Sch}{\mathcal{S}} 
\newcommand{\obs}[1]{{\langle #1 \rangle}}
\newcommand{\Ham}{\mathcal{H}}
\newcommand{\eff}{{\rm eff}}
\newcommand{\tot}{{\rm tot}}

\def\WDW{{\rm WDW\,\,}}

\def\rcr{\rho_{\rm max}}
\def\R{\mathbb{R}}
\def\pphi{p_{(\phi)}}
\def\f{\frac}
\def\lp{\ell_{\rm Pl}}
\def\ul{\underline}
\def\h{\hat}
\def\b_o{B_o}
\def\scri{\mathscr{I}}

\def\be{\nopagebreak[3]\begin{equation}}
\def\ee{\end{equation}}
\def\ba{\nopagebreak[3]\begin{eqnarray}}
\def\ea{\end{eqnarray}}

\begin{document}

$\hphantom{.}$\vspace{-2cm}
\begin{flushright}
  IGC-11/11-{5}
\end{flushright}

\title{Positive cosmological constant in loop quantum cosmology}

\author{Tomasz Paw{\l}owski${}^{1,2,3}$}
\email{tpawlows@unb.ca}

\author{Abhay Ashtekar${}^{2}$}
\email{ashtekar@gravity.psu.edu}

\affiliation{${}^{1}$Department of Mathematics and Statistics,
  University of New Brunswick, Fredericton, NB, Canada E3B 5A3.
  \\
  ${}^{2}$Institute for Gravitation and the Cosmos, Physics
  Department, Penn State, University Park, PA 16802, U.S.A.
  \\
  ${}^{3}$Department of Mathematical Methods in Physics,
  University of Warsaw, ul. Ho\.{z}a 74, 00-681 Warsaw, Poland.
}

\begin{abstract}

The k=0 Friedmann Lemaitre Robertson Walker model with a
positive cosmological constant and a massless scalar field is
analyzed in detail. If one uses the scalar field as relational
time, new features arise already in the Hamiltonian framework
of classical general relativity: In a \emph{finite} interval of
relational time, the universe expands out to infinite proper
time and zero matter density. In the deparameterized quantum
theory, the true Hamiltonian now fails to be essentially
self-adjoint both in the Wheeler DeWitt (\WDW) approach and in
LQC. Irrespective of the choice of the self-adjoint extension,
the big bang singularity persists in the \WDW theory while it
is resolved and replaced by a big bounce in loop quantum
cosmology (LQC). Furthermore, the quantum evolution is
surprisingly insensitive to the choice of the self-adjoint
extension. This may be a special case of an yet to be
discovered general property of a certain class of symmetric
operators that fail to be essentially self-adjoint.

\end{abstract}


\maketitle

\section{Introduction}
\label{s1}

Loop quantum cosmology (LQC) of the k=0, $\Lambda=0$ Friedmann
Lemaitre Robertson Walker (FLRW) model with a massless scalar field
was discussed in detail in \cite{aps3}. The scalar field serves as a
viable internal time variable both in the classical and the quantum
theory, with respect to which relational observables such as the
matter density and curvature evolve \cite{aps1,aps2}. This makes it
possible to explicitly construct the physical Hilbert space and
introduce relational Dirac observables to unravel physics of the
Planck regime in a large number of cosmological models \cite{as},
and a scheme has been sketched even for full general relativity
\cite{warsaw-full}. Using this setup it was rigorously established
that, while the big bang singularity persists in the \WDW theory of
the k=0, $\Lambda=0$ model, it is resolved due to the quantum
geometry effects of loop quantum gravity (LQG) \cite{aps3}.%
\footnote{Recently, this result has been conceptually sharpened
using the consistent histories framework in which one can
calculate probabilities for the occurrence of certain histories
without recourse to external measurement devices or interaction
with environment. Using appropriate coarse grained histories
which completely decohere, it was shown that the probability of
encountering a singularity in the distant past or future is 1
in the \WDW theory and 0 in LQC for any state (which is in the
domain of operators used to construct coarse-grained histories)
\cite{consistent2,consistent3,as}.}

An appendix in \cite{aps3} also outlined how the cosmological
constant $\Lambda$ with either sign can be incorporated. A
subsequent, detailed discussion of the $\Lambda<0$ case appeared in
\cite{bp}. It firmly established that, as in the $\Lambda=0$ case,
in LQC the big bang singularity is replaced by a quantum bounce
which occurs when the total energy density $\rho_{\rm tot}$ reaches
its maximum value $\rcr$. Furthermore the numerical value of $\rcr$
is the same as in the $\Lambda=0$ case, $\rcr \approx 0.41 \rho_{\rm
Pl}$, although now $\rho_{\rm tot}$ includes a contribution from the
cosmological constant in addition to the matter density $\rho$;\,
$\rho_{\rm tot} = \rho + \Lambda/8\pi G$. It turns out that, by a
suitable choice of time variable (or lapse function), the
$\Lambda=0$ model can be solved exactly \cite{acs}. This is not the
case for $\Lambda\not=0$. Therefore, results of \cite{bp} for the
$\Lambda <0$ case are conceptually important also because they
demonstrate that the LQC bounce and the qualitative features of the
resulting Planck scale physics are not tied to exact solvability.
Finally, although the situation with the bounce is the same, the
presence of the cosmological constant does alter the underlying
mathematical structure in non-trivial ways. In particular, in the
deparameterized picture, while the spectrum of the true Hamiltonian
is continuous in the $\Lambda=0$ case, it is purely discrete in the
$\Lambda <0$ case.

The goal of this paper is to present an analogous, detailed account
of the $\Lambda >0$ case. Even though we will again consider a
massless scalar field, rather surprisingly, the flip of the sign of
the cosmological constant changes the underlying mathematical and
conceptual structure significantly. Let us begin with the classical
theory. If one again uses the scalar field $\phi$ for internal time,
in contrast to the $\Lambda =0$ and $\Lambda <0$ cases
\cite{aps3,bp}, the Hamiltonian vector field on the phase space is
now incomplete. As a result, volume of any compact co-moving region
becomes infinite and the matter density vanishes at a finite instant
$\phi_o$ of internal time $\phi$. This situation is qualitatively
similar to that in the case of a non-relativistic particle in a
steep negative potential whose dynamical trajectories reach infinity
in a finite time. In such situations, typically, the Hamiltonian
operator in Schr\"odinger quantum mechanics is symmetric but not
essentially self-adjoint. Each self-adjoint extension then yields a
unitary evolution but evolutions obtained from distinct operators
are both mathematically and physically inequivalent. In the present
case, one again finds that the true Hamiltonian operator generating
evolution in the scalar field time is symmetric but not essentially
self-adjoint. However, rather surprisingly, this ambiguity has
negligible effect on states of physical interest: those that start
out being peaked at a classical solution in a low curvature region.
In particular, all these states undergo a quantum bounce and the
total density $\rho_{\rm tot}$ at the bounce is again universal.
Furthermore, while the evolution of expectation values of physical
observables does depend on the choice of self-adjoint extension, the
dependence is extremely weak. This robustness may be related to the
fact that, on the classical phase space, one can extend both the
evolution equations and the solutions simply by analytical
continuation, without having to introduce specific boundary
conditions at infinity. Our analysis raises the possibility that
there may well be a general pattern and new results could be found
on properties of certain sub-classes of operators that fail to be
essentially self-adjoint.

The paper is organized as follows. In Sec. \ref{s2} we discuss
the Hamiltonian framework for the k=0, $\Lambda>0$ model. Sec.
\ref{s3} is devoted to the \WDW quantum theory and Sec.
\ref{s4} to LQC. We conclude in \ref{s5} with a brief summary
and discussion. Because the numerical simulations in this
paper were completed soon after the initial analysis in
\cite{aps3}, they use an older value of LQC area gap which
turned out to be half the value that is relevant for states
used in LQC \cite{awe1}. In the main text we use this more
recent value. Therefore, unfortunately, in Sec. \ref{s4} there
is an occasional mismatch of factors of two between the text
and the figures.

\section{Hamiltonian Framework}
\label{s2}

Because our primary focus is on LQC, in this section we will
summarize the phase space formulation of the FLRW model under
consideration in terms of variables that descend from LQG. (For
details, see, e.g. \cite{as}). The space-time manifolds
$\mathcal{M}$ will be taken to be topologically $\R^4$, equipped
with a preferred foliation by spatially homogeneous and isotropic
space-like slices $M$. The space-time metrics $g$ will have the form
\begin{equation}
g= -N^2 \rd t^2 + a^2(t)\, q^o
\end{equation}
where $q^o$ is the fixed, positive definite, flat metric on $M$
(determined by the co-moving coordinates), $N$ is the lapse function
and $a(t)$ the scale factor. Since all physical fields in the model
are spatially homogeneous, and since $M$ is non-compact, integrals
representing the symplectic structure and Hamiltonians trivially
diverge. One therefore introduces an infrared regulator ---a cell
$\mathcal{C}$ taken to be cubical with sides along co-moving
coordinates--- and restricts all integrals to it. The phase space
structure and intermediate results depend on the choice of
$\mathcal{C}$ and $q^o$. Therefore one has to either show that the
\emph{final} physical results are independent of these choices or
remove the infrared regulator by letting $\mathcal{C}$ to expand out
to fill $M$.

In LQG, one uses triads in place of 3-metrics. The freedom in the
choice of their orientation enables one to introduce a configuration
variable $v$ which captures both the volume $V$ of the cell
$\mathcal{C}$ determined by $q$ and the orientation of the
underlying physical triad:
\be ({\rm sgn}\, v) \,\, v = \f{V}{2\pi\gamma\lambda\lp^2} \equiv
\f{a^3V_o}{2\pi\gamma\lambda\lp^2} \ee
where $\gamma$ is the Barbero-Immirzi parameter of LQG, $\lambda^2 =
4\pi\sqrt{3}\gamma\, \lp^2$ is the `area gap' of LQG that is
relevant to LQC \cite{awe1} and $V_o$ the volume of $\mathcal{C}$
with respect to $q^o$. ($\rm{sgn}\, v$ is positive if the physical
triad is oriented along the fiducial triad compatible with $q^o$ and
negative if the orientations are opposite.) The canonically
conjugate momentum is denoted by $b$. On classical solutions, it is
given by
\be b = \gamma \lambda H \, \equiv \, \gamma\lambda \f{1}{a}\f{\rd
a}{\rd t} \ee
where $H$ is the Hubble parameter and $t$ is the proper (or
cosmological) time. In these definitions, factors involving $\gamma,
\lambda$ and $\lp$ are introduced to simplify the final expression
of the Hamiltonian constraint operator later
on.%
\footnote{The variable $v$ is the same as that used in
\cite{aps3} and is related to the variable $\nu$ of
\cite{acs,as} via: $\nu = \lambda v$ and the variable $b$ is
related to ${\rm b}$ used in \cite{acs,as} via ${\rm b} =
(1/\lambda) b$. These relative rescalings by $\lambda$ make the
variables $v,b$ used here dimensionless which renders a
considerable algebraic simplification in various expressions
and equations.}
For the scalar field, the basic canonical pair is, as usual, $\phi,
\pphi$. Thus the phase space is topologically $\mathbb{R}^4$ and
equipped with basic Poisson brackets:
\be \{b,\, v\} = \f{2}{\hbar}, \quad {\rm and}\quad \{\phi,
\pphi\} =1\, . \ee
Because of a gauge fixing tailored to spatial homogeneity and
isotropy, the Gauss and the diffeomorphism constraints of LQG
are automatically satisfied \cite{abl}. Thus, we are left only
with the Hamiltonian constraint. Since we want to use the
scalar field $\phi$ as internal time and since $\phi$ satisfies
the wave equation on $(\mathcal{M}, g)$, it is appropriate to
use the lapse field $N$ that is adapted to a harmonic time
coordinate $\tau$, satisfying $\Box \tau=0$ \cite{as}. As in
\cite{acs}, this is achieved by setting $N := a^3$. Then the
Hamiltonian constraint is given by:
\begin{equation}\label{eq:Cgr-wdw}
{C} = \, \pphi^2 - 3\pi\hbar^2 G b^2v^2 +
\pi \gamma^2\lambda^2\, \hbar^2 G\, \Lambda\, v^2\,  \approx\, 0 ,
\end{equation}
Note that while $b\in (-\infty, \infty)$, on the constraint surface
it must satisfy $|b| \ge b_o = \gamma\lambda\sqrt{\Lambda/3}$. The
equations of motion for the scalar field, generated by the
Hamiltonian constraint, are:
\begin{equation}\label{eq:EOM-matt}
  \f{\rd{p}_{(\phi)}}{\rd \tau}= 0  \quad {\rm and} \quad
  \f{\rd{\phi}}{\rd \tau} = 2\pphi ,
\end{equation}
Eq. \eqref{eq:EOM-matt} implies that $\pphi$ is a constant in
any solution and $\phi(\tau) = 2\pphi \tau +{\rm const} $.
Therefore, in any space-time defined by a phase space dynamical
trajectory, $\phi$ can be used as an evolution parameter in
place of $\tau$. This is our relational time variable. The
equation of motion for $v$ in the relational time $\phi$ is
given by
\begin{equation}\label{eq:Fried}
  [\partial_{\phi}v]^2 = 12\pi G\, [v^2\, +\,
  \f{\pi\gamma^2\lambda^2\lp^2\hbar\Lambda}{2\pphi^2}\, v^4 ] \, .
\end{equation}
and its solution, expressed in terms of the physical volume $V =
(2\pi\gamma\lambda\lp^2)\, |v|$ of the cell $\mathcal{C}$ is:
\be\label{eq:v-phi}
 V(\phi) =  \big(\f{\sqrt{4\pi G} \, \pphi}{\sqrt{\Lambda}}\big) \,\,
 \f{1}{|{\rm sinh}[\sqrt{12\pi G}\,(\phi-\phi_o)]|} \ee
where $\phi_o$ is a constant (that can vary from solution to
solution). Eq. \eqref{eq:v-phi} implies that for each $\phi_o$
we have two types of solutions: Those that start at
$\phi=-\infty$ with a big bang singularity, i.e. with zero
volume for the cell $\mathcal{C}$, and expand out to infinite
volume at $\phi =\phi_o$, and those that start out at infinite
volume at $\phi=\phi_o$ and contract into a big crunch
singularity at $\phi=\infty$ where $\mathcal{C}$ shrinks to
zero volume. In either case, the evolution ends at the finite
value $\phi_o$ of the relational time $\phi$. Put differently,
the Hamiltonian vector field generating evolution in $\phi$ is
incomplete.

An obvious question then is: Can the phase space evolution be
naturally extended beyond $\phi=\phi_o$? We will now show that a
mathematically natural extension does exist. Consider first matter
density $\rho = \pphi^2/2V^2$ of the scalar field, which is a
physical observable of direct physical interest. Its time dependence
is given by
\begin{equation}
  \rho(\phi) = \frac{\Lambda}{8\pi G}\,
  \sinh^2[\sqrt{12\pi G}(\phi-\phi_o)]\, .\end{equation}
Since it is analytic in $\phi$, in the $\rho$-$\phi$ plane the
dynamical trajectory represented by the `contracting branch' is
simply an analytical continuation of `expanding branch'. In
space-time terms the full phase space trajectory can be interpreted
as follows: the universe starts out with a big bang at
$\phi=-\infty$, expands out till the matter density $\rho$ becomes
zero at $\phi = \phi_o$, and then starts contracting, ending in a
big crunch at $\phi= \infty$. Situation is similar with respect to
$b$. The dynamical trajectory in the $b$-$\phi$ plane also extends
analytically from the expanding to the contracting branch:
\begin{equation}
  b(\phi) = \pm\, b_o\,\cosh[\sqrt{12\pi G}(\phi-\phi_o)]\quad
  {\rm where}\quad  b_o:=\gamma\lambda\, \sqrt{\Lambda/3}\, .
\end{equation}
Indeed, in both these cases, it seems artificial to stop the
evolution of the expanding branch at $\phi=\phi_o$ and say that the
contracting branch, $\phi \ge \phi_o$, is a distinct trajectory.

In the space-time picture, on the other hand, as $\phi$ approaches
$\phi_o$, the proper time goes to $\infty$ in the expanding branch
and $-\infty$ in the contracting branch. Therefore the space-time
represented by the expanding branch is future complete and that
represented by the contracting branch is past complete. In effect,
the extended space-time can be obtained by gluing together future
null infinity of the expanding branch with the past null infinity of
the contacting branch. From the perspective of space-time geometry,
this gluing is just an optional mathematical construct. From the
perspective of the Hamiltonian framework based on the relational
time, on the other hand, the extension is natural and even appears
to be necessary to have a complete picture of evolution. Since the
Hamiltonian framework can be regarded as the imprint left on the
classical regime by the quantum theory, one may suspect that the
extension may have its true origin in the mathematical and
conceptual framework underlying quantum cosmology. In the next two
sections we will see that this is indeed the case.

\section{The Wheeler-DeWitt Theory}
\label{s3}

In this section we will first introduce the kinematical
structure of the \WDW quantum theory, then investigate
properties of the operator $\Theta_\Lambda$ representing the
gravitational part of the Hamiltonian constraint and finally
discuss dynamics. As mentioned in section \ref{s1}, because of
the presence of a positive cosmological constant,
$\Theta_\Lambda$ fails to be essentially self-adjoint. Much of
the discussion is devoted to establishing this property and
exploring its consequences.

\subsection{Quantum kinematics}
\label{s3.1} \label{sec:wdw-rep}

In the Dirac program of quantization of constrained systems, one
first ignores the constraints and constructs a kinematical Hilbert
space $\Hil_{\kin}$, the quantum analog of the full phase space of
the classical theory. The quantum constraint is then written as an
operator on $\Hil_{\kin}$. Physical states lie in the kernel of this
operator.

In the \WDW theory, one takes the kinematic Hilbert space to be
$\ub{\Hil}_{\kin}= \ub{\Hil}_{\gr} \otimes \ub{\Hil}_{\phi}$, where,
as is common in the LQC literature \cite{as}, under-bars emphasize
that the symbols refer to the \WDW theory. As in the textbook
Schr\"odinger quantum mechanics, one sets $\ub{\Hil}_{\phi} =
L^2(\re,\rd \phi)$ and $\ub{\Hil}_{\gr} := L^2_S(\re,\rd v)$. Here
the subscript $S$ denotes that the states are symmetric, i.e.,
satisfy $\ub{\psi}(v) = \ub{\psi}(-v)$, thereby encoding the fact
that since  $v\to -v$ results from an orientation flip of the
physical triad $E^a_i$, it is a large gauge transformation
\cite{aps2} under which physics of the model does not change. For
later purposes we note that the inner product between two states
$\ub{\psi},\ub{\chi}$ on the gravitational Hilbert space
$\ub{\Hil}_{\gr}$ is given simply by
\begin{equation}\label{eq:gr-ip-wdw}
\sip{\ub{\psi}}{\ub{\chi}} = \int_{\re} \rd v\, {\ub{\bar\psi}}(v)
\ub{\chi}(v) .
\end{equation}

It is straightforward to write down the quantum operator
corresponding to the constraint function \eqref{eq:Cgr-wdw}:
\ba \label{eq:C-wdw} \ub{\hat{C}} &=& \id\otimes\partial_{\phi}^2 +
\ub{\Theta}_{\Lambda} \otimes \id \qquad {\rm with}\\
\ub{\Theta}_{\Lambda} := \ub{\Theta}_o -\pi G \gamma^2\lambda^2
\Lambda v^2\, \id \,\,\, &&{\rm and}\quad \ub{\Theta}_o = - 12\pi
G\sqrt{|v|}\partial_v|v|\partial_v\sqrt{|v|}\, ,  \ea
where in the expression of $\ub{\Theta}_o$  we have chosen a factor
ordering that is compatible with the one used in LQC in section
\ref{s4} (see \cite{aps3} for details).

Physical states lie in the kernel of $\ub{\hat{C}}$. To find
this kernel and endow it with a Hilbert space structure, it is
natural to use the general `group averaging method'
\cite{dm,almmt,abc} as in \cite{aps2,klp1}. The implementation
requires that the operator $\ub{\Theta}_{\Lambda}$ be
self-adjoint and uses its spectral decomposition. In the
$\Lambda\le 0$ cases these steps could be readily carried out
and the physical sector of the theory could be constructed in a
rather straightforward manner \cite{aps3,bp}. With a positive
cosmological constant, on the other hand, it turns out that the
eigenfunctions $\ub\psi(v)$ of $\ub{\Theta}_{\Lambda}$ (with
real eigenvalues) are Bessel functions of imaginary order.
Unfortunately, for our purposes, they are difficult to work
with (see for example \cite{AbramowitzStegun}, for their
properties). Therefore, it turns out to be more convenient to
pass to the dual, $b$-representation and work with the
wave-functions $\ul\psi(b)$.

Let $\Dom$ denote the dense domain in $L^2(\re,\rd v)$ consisting of
smooth functions which, together with all their derivatives, fall
off faster than any polynomial at infinity (i.e. let $\Dom$ be the
Schwartz space). The operator $\ub\Theta_\Lambda$ is symmetric on
$\Dom$. Given any state $\ub\psi(v) \in \Dom$, the corresponding
wave function in the $b$ representation can be obtained by a Fourier
transform. We will set
\begin{equation}\label{eq:v-b-trans}
  [\ub{\Fou}\psi](b) = \frac{1}{2\sqrt{\pi}}
  \int_{\re} \rd v\, |v|^{-\f{1}{2}}\, \psi(v)\, e^{\f{i}{2}\,vb} ,
\end{equation}
so that, using the fact that $\h{v}\ub\psi(b) = - 2i \partial_b
\ub\psi(b)$ in the $b$ representation, the operator
$\ub{\Theta}_{\Lambda}$ assumes a convenient form:
\begin{equation} \label{gravcon}
  \ub{\Theta}_{\Lambda} = - 12\pi G [\, (b\,\partial_b)^2 - b_o^2 \partial_b^2\, ]
  \, .
\end{equation}
where $b_o=  \gamma\lambda\,\sqrt{\Lambda/3}$. We will first discuss
certain properties of $\ub{\Theta}_{\Lambda}$ and then use them to
construct the physical sector of the theory.

\subsection{Properties of $\Theta_\Lambda$}
\label{s3.2}

\subsubsection{Weak solutions to the eigenvalue equation}

To discuss dynamics, we have to extend $\Theta_{\Lambda}$ to a
self-adjoint operator. To analyze existence and uniqueness of these
extensions, we need weak solutions to the eigenvalue problem
${\ub\Theta}_\Lambda {\ub\psi}_{\zeta} = {\zeta} {\ub\psi}_{\zeta}$,
i.e., distributions ${\ub\psi}_{\zeta}$ such that
\begin{equation}\label{eq:eigenf}
(\ub{\psi}_{{\zeta}}|\ub{\Theta}_{\Lambda}^{\dagger}-
  \bar{{\zeta}}\id\,\ket{\ub{\chi}} = 0 \qquad\qquad \forall \ub{\chi}\in\Dom \ ,
\end{equation}
where, as usual, the action $(\ub{\psi}_{{\zeta}}\ket{\ub{\chi}}$ of
the distribution $(\ub{\psi}_{\zeta}|$ on the test function
$\ket{\ub{\chi}}$ is defined using the Hilbert space inner product
on $\Hil_{\gr}$.

Now, because of the factor $|v|^{1/2}$ ---introduced to simplify the
form of the constraint operator--- the form of the inner product in
the $b$ representation is not transparent. To spell it out, let us
first note that the action of $\Theta_\Lambda$
leaves subspaces containing wave functions $\ub\chi$ with support on
positive and negative $v$-axis separately invariant. Therefore, we
divide the solutions $\ub{\psi}_{{\zeta}}$ to \eqref{eq:eigenf} into
parts $\ub{\psi}_{{\zeta}}^{\pm}$
\begin{equation}
  \ub{\psi}_{{\zeta}} \mapsto \ub{\psi}_{{\zeta}}^{\pm} = \theta(\pm v)
\, \ub{\psi}_{{\zeta}}
\end{equation}
by considering the inner product of $\psi_{\zeta}$ with test
functions which have support only on the positive or negative parts
of the $v$-line. Since by \eqref{eq:gr-ip-wdw} the states supported
on $\re^+$ are always orthogonal to the ones supported on $\re^-$
the action of $\ub{\psi}_\zeta$ can be written as
\begin{equation}\label{eq:ip-comp}\begin{split}
  (\ub{\psi}_{{\zeta}}\ket{\ub{\chi}}
  &= (\ub{\psi}_{{\zeta}}^{+}\ket{\ub{\chi}}_{+} +
  (\ub{\psi}_{{\zeta}}^{-}\ket{\ub{\chi}}_{-},\qquad{\rm where} \\
  (\ub{\psi}_{{\zeta}}^{\pm}\ket{\ub{\chi}}_{\pm}
  &:= 4\int_{\re}\rd b\, \bar{\ub{\psi}}^{\pm}_{{\zeta}}(b)\,\,
  [\pm i\partial_b] \ub{\chi}(b)\qquad \forall \ub{\chi}\in\Dom\, .
\end{split}\end{equation}

With this explicit form of the action of $\ub{\psi}_\zeta$ at hand,
we can now solve \eqref{eq:eigenf}. For this, it is convenient to
introduce a change of variables to simplify the form of
$\ub{\Theta}_{\Lambda}$. Let us set
\begin{equation}\label{eq:wdw-x-def}
  x := \begin{cases}
         \arctan(b/\sqrt{|b^2-b_o^2|}) , & |b|<b_o , \\
         \begin{aligned}
           \sgn(b)[\pi/2 - \ln(b_o) +\ln(|b|+\sqrt{|b^2-b_o^2|})]
         \end{aligned},
          & |b|>b_o .
       \end{cases}
\end{equation}
or, reciprocally
\be \label{eq:wdw-b-def}
         b = \begin{cases}
         b_o\, \sin x  , & |x|< \pi/2 , \\
        \begin{aligned}
         b_o\, {\rm cosh}(|x| - \f{\pi}{2})\, {\rm sgn}(x)
         \end{aligned},
        & |x| > \pi/2 .
\end{cases}
\end{equation}
where, as before, $b_o:=\gamma\lambda\, \sqrt{\Lambda/3}$. (Recall
from section \ref{s2} that, in the classical solutions, $|v| \to
\infty$ as $b \to \pm b_o$, or equivalently, as $x \to \pm \pi/2$.)
Then, except
at points $b=\pm b_o$ we have%
\footnote{While $\Theta_{\Lambda}$ preserves the space of
  smooth functions of $b$, on functions of $x$ its action is
  discontinuous at $x=\pm \pi/2$ because, although $b$ is a smooth
  function of $x$,\, $\rd b/\rd x =0$ there. Since we are looking for
  distributional solutions, this discontinuity is harmless.
}
\begin{equation}\label{eq:theta-wdw-x}
  \ub{\Theta}_{\Lambda} = -12\pi G \,\sgn(|x|-\pi/2)\,\,\partial_x^2\, .
\end{equation}
Since $\ub{\Theta}_\Lambda$ is just proportional to the simple
operator $\partial_x^2$ except at $x =\pm \pi/2$, and we are
interested in distributional eigenfunctions which are symmetric in
$x$, they are necessarily of the form
\begin{equation}\label{eq:eig-sol}
  \ub{\psi}_{{\zeta}}(x)\,=\,\,
  \begin{cases}
    A_+ e^{i\sqrt{\zeta}x}  + A_- e^{-i\sqrt{\zeta}x}, & x > \f{\pi}{2}   \\
    B (e^{\sqrt{\zeta}x}  +  e^{-\sqrt{\zeta}x}), & |x| < \f{\pi}{2}      \\
    A_+ e^{-i\sqrt{\zeta}x}  + A_- e^{i\sqrt{\zeta}x}, & x < \f{\pi}{2}
  \end{cases}
\end{equation}
for constants $A_\pm, B_\pm$ that satisfy suitable `gluing
conditions'.

To determine these conditions we use the fact that the components
$\ub{\psi}_{{\zeta}}^{\pm}$ are independent solutions to
\eqref{eq:eigenf} and apply the decomposition \eqref{eq:eig-sol} of
them directly to \eqref{eq:eigenf}, using the form
\eqref{eq:ip-comp} of the inner product. Splitting the domain of
integration of \eqref{eq:ip-comp} onto three intervals
$\mathcal{I}_i\in\{\,\, ]-\infty,-\pi/2], [-\pi/2,\pi/2],
[\pi/2,\infty[\,\, \}$ and integrating the resulting expression
twice by parts we obtain
\begin{subequations}\label{eq:eig-ip-parts}\begin{align}
  (\ub{\psi}_{{\zeta}}^{\pm}|\ub{\Theta}_{\Lambda}^{\dagger}-
  \bar{{\zeta}}\id\ket{\ub{\chi}}_{\pm} 
  = &\mp 4i \hspace{-0.5cm} \sum_{s\in\pm 1,
  \sigma\in\{+,-\}} \hspace{-0.3cm}
    \sigma s \lim_{x\to^{\sigma}(s\pi/2)}
    \bar{\ub{\psi}}_{{\zeta}}^{\pm}(x)
    [\ub{\Theta}_{\Lambda}\ub{\chi}](x) \notag \\
  &\mp 48i\pi G \hspace{-0.5cm} \sum_{s\in\pm 1,
  \sigma\in\{+,-\}} \hspace{-0.3cm}
    \sigma s \lim_{x\to^{\sigma}(s\pi/2)}
    \sgn(|x|-\pi/2) [\partial_x\bar{\ub{\psi}}_{{\zeta}}^{\pm}](x)
    [\partial_x\ub{\chi}](x) \notag \\
  &\pm 4i \sum_{i=1}^3 \int_{\mathcal{I}_i} \rd x\,
    (\partial_x\ub{\chi})\,\,[\ub{\Theta}_{\Lambda}-\bar{{\zeta}}]
    \bar{\ub{\psi}}_{{\zeta}}^{\pm}
    \tag{\ref{eq:eig-ip-parts}}\, ,
\end{align}\end{subequations}
where $x\to^{\sigma}\!(s\pi/2)$ denotes the limit as $x$ approaches
$s\pi/2$ from above if $\sigma = +$ and below if $\sigma = -$. Now,
the integrand of the third term on the right hand side vanishes
identically because $\ub{\psi}_{\zeta}^\pm$ are given by
(\ref{eq:eig-sol}), and the second term on the right side also
vanishes because smoothness (in $b$) of $\ub{\chi}$ implies that
$\partial_x\ub{\chi}$ at $x=\pm \pi/2$. Therefore, only nontrivial
contributions to the right side of (\ref{eq:eig-ip-parts}) come from
the first term. Since $\ub{\Theta}_{\Lambda}\ub{\chi}$ does not
generically vanish at $x=\pm \pi/2$ we conclude that \emph{weak
solutions to the eigenvalue problem are given by (\ref{eq:eig-sol})
where the coefficients are chosen so that
$\ub{\psi}_{{\zeta}}=\ub{\psi}_{{\zeta}}^{+}+
\ub{\psi}_{{\zeta}}^{-}$ is continuous in $x=\pm\pi/2$} (but not
necessarily differentiable).

\subsubsection{Self-adjoint extensions of $\ub{\Theta}_{\Lambda}$}

\textbf{Deficiency Spaces:}\,\, The operator $\ub{\Theta}_\Lambda$
is symmetric on $\ub{\Hil}_{\gr}$ and the operator $\ub{\Theta}_o$
is known to be essentially self-adjoint \cite{warsaw3}. However, the
cosmological constant term acts like a negative unbounded potential.
Therefore, from one's experience with Hamiltonians in
non-relativistic quantum mechanics, one would not expect the
operator $\ub{\Theta}_{\Lambda}$ to be essentially self-adjoint.
Indeed, its LQC counterpart was recently shown to admit a family of
inequivalent extensions \cite{kp-posL}. We will now show that this
general expectation is correct by analyzing \emph{deficiency spaces}
\cite{ReedSimon-v2} of $\ub{\Theta}_{\Lambda}$. For notational
simplicity, we will first rescale this operator and consider
$\ub{\Theta}'_{\Lambda}= (12\pi G)^{-1}\, \ub{\Theta}_{\Lambda}$.

The deficiency spaces $\defi^{\pm}$ are spanned by (kinematically)
normalizable solutions $\ub{\psi}^{\pm}$ to equation
\eqref{eq:eigenf} with the eigenvalue ${\zeta}=\pm 8i$
\footnote{The deficiency spaces as defined in \cite{ReedSimon-v2}
correspond to the eigenvalues ${\zeta}=\pm i$, however one can equally
work with the spaces corresponding to ${\zeta}=\pm ir$ where $r$ is any
positive real number. We chose $r=8$ just to simplify notation in
subsequent calculations}.
Their elements are solutions to \eqref{eq:theta-wdw-x} which are
symmetric and everywhere continuous. Therefore, from
(\ref{eq:eig-sol}) it follows that they take the general form
\begin{equation}\label{eq:defi-wdw-x-ext}
  \ub{\psi}^{\pm}(x) = A\, e^{2(1\mp i)|x|}\, +\, B\, e^{-2(1\mp i)|x|} ,
\end{equation}
for $|x|>\pi/2$, where $A,B$ are some constants. However, we will
now show that only the solutions with $A=0$ are normalizable.
Heuristically this is plausible because the first term on the right
hand side diverges as $|x| \to \infty$ while the second goes to zero
in this limit. But to establish the result we need to consider the
Hilbert space norms. Since the inner product is simple in the $v$
representation, let us solve the equation directly in that
representation. Using $\ub{\Theta}_{\Lambda}$ from \eqref{eq:C-wdw}
we can express $\ub{\psi}^{\pm}$ as a linear combination of Bessel
functions
\begin{equation}\label{eq:defi-wdw-v}
  \ub{\psi}^{\pm}(v)
  = C\, \f{Y_{\pm 2(1-i)}(-ic|v|)}{\sqrt{|v|}} + D\,
\f{J_{\pm 2(1-i)}(-ic|v|)}{\sqrt{|v|}}\, .
\end{equation}
Since these functions decay at infinity sufficiently fast, to find
if they are normalizable it suffices to focus on their behavior for
small $v$. In this limit, the solution approaches
\begin{equation}
  \ub{\psi}^{\pm}(v) \to E_{-}\, \f{e^{-2(1\mp i)\ln|v|}}{\sqrt{|v|}}
  + E_{+}\, \f{e^{2(1\mp i)\ln|v|}}{\sqrt{|v|}}\, ,
\end{equation}
whence from the form of the inner product \eqref{eq:gr-ip-wdw} it
follows that only the solution with $E_{-}=0$ is normalizable.
Since this normalizable solution goes as $|v|^{3/2}$, its Fourier
transform \eqref{eq:v-b-trans} is well-defined and square integrable
in $b$. Having established this property, we can now return to the
eigenfunctions \eqref{eq:defi-wdw-x-ext} in the $x$ representation.
Since ${\rd} b/{\rd} x \sim {\rm sinh}(x)$, the normalizable
solution $\ub{\psi}(x)$ must have the property that $\int {\rd}x\,
{\rm sinh}(x)\, |\ub{\psi}|^2 < \infty$. This condition implies that
the normalizable solution is given by \eqref{eq:defi-wdw-x-ext} with
$A=0$.

Setting $A=0$ and using the symmetry properties of elements of
$\ub{\Hil}_{\gr}$ and the continuity property of eigenfunctions we
conclude that elements of deficiency subspaces $\defi^{\pm}$ have
the form
\begin{equation}
  \ub{\psi}^{\pm}(x) = B \begin{cases}
  \frac{1}{2}[e^{2(1\pm i)x} + e^{-2(1\pm i)x} ], & |x|\leq \f{\pi}{2} \\
  \cosh(\pi)\,\, e^{-(2(1\mp i)|x|-\pi)} , & |x|\geq\f{\pi}{2} \end{cases}
\end{equation}
for some $B\in\compl$. From this it follows immediately that
$\defi^{\pm}$ are $1$-dimensional.\\

\textbf{Self-Adjoint Extensions and their Domains:}\,\,\, The fact
that $\defi^\pm$ are non-empty immediately implies that
$\ub{\Theta}'_{\Lambda}$ admits a family of inequivalent
self-adjoint extensions \cite{ReedSimon-v2}. Elements of this family
are labeled by the unitary transformations $U:\defi^+\to\defi^-$. In
our case they are all of the form
\begin{equation}\label{eq:wdw-ext-trans}
  U_{\alpha}:\ub{\psi}_o^{+} \mapsto e^{i\alpha} \ub{\psi}_o^- ,
\end{equation}
where $\ub{\psi}_o^{\pm}$ are some chosen \emph{normalized} elements
of $\defi^{\pm}$. Thus, $\ub{\Theta}'_{\Lambda}$ admits a
1-parameter family of self-adjoint extensions, labeled by $\alpha
\in [0,2\pi)$. Using theorem X.2 of \cite{ReedSimon-v2}, we conclude
that the domains $\Dom_{\alpha}$ of these extensions are given by
\begin{equation}\label{eq:dom-ext}
\Dom_{\alpha} = \{ \ub{\psi} + \ub{\psi}^{+} +
U^{\alpha}\ub{\psi}^{+} ;\
\ub{\psi}\in\Dom,\, \ub{\psi}^{\pm}\in\defi^{\pm} \}\, .
\end{equation}
Up to a constant rescaling, the terms $\psi_{\alpha}:=
\ub{\psi}^{+} + U^{\alpha}\ub{\psi}^{+}\in\Dom_{\alpha}$ that
depend on and characterize the extension are given by
\begin{equation}\label{eq:wdw-ext-char}
  \ub{\psi}_{\alpha}(x) = \begin{cases}
  \begin{aligned}
  (1/2)\, [ &e^{2x}\cos(2x-\alpha/2)
  + e^{-2x}\cos(2x+\alpha/2) ]
  \end{aligned} \ , & |x|\leq\pi/2 ,\\
  \cosh(\pi) e^{-(2|x|-\pi)}\cos(2|x|-\alpha/2) , & |x|\geq\pi/2\, .
  \end{cases}
\end{equation}
Although these functions are continuous, generically, they are not
differentiable at $x=\pm\pi/2$. On the other hand, each element
$\ub{\psi}$ of $\Dom$ is differentiable in $b$ and satisfies
$[\partial_x \ub{\psi}](x=\pm\pi/2)=0$. Therefore, for fixed value
of $\alpha$ the ratio between the left and right hand derivative of
any element of $\Dom_{\alpha}$ is a constant, common for all
$\Dom_{\alpha}$, depending only on $\alpha$. Furthermore, the
function
\begin{equation}\label{eq:wdw-ext-cond}
\beta(\alpha) :=
\arctan\left( \frac{[\partial_x^- \ub{\psi}_{\alpha}](x=\pi/2)}%
{[\partial_x^+ \ub{\psi}_{\alpha}](x=\pi/2)} \right)\,\, \in [0,\pi) ,
\end{equation}
(where $\partial_x^\pm$ denote derivatives from above and below
respectively) is monotonic:
\begin{equation}
  [\partial_{\alpha}\beta](\alpha)
  = \frac{\cosh(\pi)\sinh(\pi)}{\cosh(2\pi)-\sin(\alpha)}
  \,\,\, \in [0.498,\,\,0.502] .
\end{equation}
Therefore the relation
\begin{equation}
  [0,2\pi) \ni \alpha \mapsto \beta(\alpha) \in [0,\pi)
\end{equation}
is a bijection. Thus $\beta$ provides an alternate labeling of the
extensions which is much more convenient as it has a direct
interpretation in terms of the discontinuity in the derivatives
$\partial_x^\pm\, \ub{\psi}_{\alpha}$ at $x=\pm\pi/2$. Therefore,
labeling the extension by $\beta$ allows us to easily identify  the
eigenfunctions of $\ub{\Theta}_{\Lambda}$ which span particular
extended domain.

To summarize, the symmetric operator $\ub{\Theta}_{\Lambda}$ defined
on the domain $\Dom$ admits a 1-parameter family of self-adjoint
extensions $\ub{\Theta}_{\Lambda,\beta}$, parameterized by $\beta\in
[0,\pi)$ with domains given by \eqref{eq:wdw-ext-cond}. The
parameter $\beta$ directly captures the boundary conditions at $x =
pi/2$ satisfied by states in the domain $\Dom_\beta$ of
$\ub{\Theta}_{\Lambda,\beta}$ via \eqref{eq:wdw-ext-cond}.\\

\textbf{Eigenfunctions:} Given a self-adjoint extension
$\ub{\Theta}_{\Lambda,\beta}$  we can construct the corresponding
physical Hilbert space $\ub{\Hil}^{\phy}_\beta$ by group averaging
and discuss quantum dynamics. Both these tasks require us to find
the eigenfunctions of $\ub{\Theta}_{\Lambda,\beta}$. For this, we
first note that, since the operator $\partial_{\phi}^2$ in
\eqref{eq:C-wdw} is negative definite, only the positive part of
$\ub{\Theta}_{\Lambda,\beta}$ is relevant for the construction of
$\ub{\Hil}^{\phy}_{\beta}$. The general, symmetric eigenspaces of
$\ub{\Theta}'_{\Lambda}$ corresponding to the eigenvalue $k^2$ are
$2$-dimensional and are spanned by functions $\ub{\psi}_k$ (where
$k\in\re$) of the general form
\begin{equation}
  \ub{\psi}_k(x) = A\, \begin{cases}
  \cosh(kx) , & |x|<\pi/2 , \\
  \cosh(k\pi/2)e^{ik(|x|-\pi/2)} , & |x|>\pi/2
  \end{cases}
\end{equation}
where $A \in {\mathbb{C}}$. The eigenfunctions spanning the
particular extended domain $\Dom_{\beta}$ need to satisfy the
condition analogous to \eqref{eq:wdw-ext-cond}. They are therefore
given by
\begin{equation}\label{eq:wdw-ext-eigenf}
  \ub{\psi}_{\beta, k}(x) = B\, \begin{cases}
  \frac{\cos(k\pi/2+\sigma_{\beta}(k))}{\cosh(k\pi/2)} \cosh(kx) ,
  & |x|<\pi/2 , \\
  \cos(k|x|+\sigma_{\beta}(k)) , & |x|>\pi/2 ,
  \end{cases}
\end{equation}
where $B\in\re^+$, $k\in\re^+$ and the phase shift
$\sigma_{\beta}(k)$ satisfies the relation
\begin{equation}\label{eq:wdw-ext-rot}
  \tan(k\pi/2+\sigma_{\beta}(k)) = \f{\tanh(k\pi/2)}{\tan(\beta)}\, .
\end{equation}
To find the normalization factor $B$ we use an analog of the
method used in appendix~A2 of \cite{kp-scatter}. 
Specifically, first note that, for large $|x|$,\, $\ub{\psi}_{\beta,
k}$ approaches an eigenfunction of $\ub{\Theta}'_{\Lambda=0}$ with a
controlled rate of convergence
\begin{equation}\label{eq:wdw-e-limit}\begin{split}
\ub{\psi}_{\beta, k}(x) &= \ub{\psi}^o_{\beta, k}(x)
+ O(e^{-2|\tilde{x}|}) , \\
\ub{\psi}^o_{\beta, k}(x) &:=
B\, \cos[\,k(\ln(2)-\ln(b_o)+\f{\pi}{2}+\tilde{x})+\sigma_{\beta}(k)\,] .
\end{split}\end{equation}
where $\tilde{x} = \ln|b|$. Similarly, in the $v$-representation, we
have \cite{AbramowitzStegun}
\begin{equation}
  \ub{\psi}_{\beta, k}(v) = \ub{\psi}^o_{\beta, k}(v) + O(v^2) .
\end{equation}
which implies \cite{kp-scatter} that $\ub{\psi}_{\beta, k}$ and
$\ub{\psi}^o_{\beta, k}$ satisfy the same (distributional)
normalization conditions. Therefore, the known normalization of
$\ub{\psi}^o_{\beta, k}$ \cite{acs} and the relation between
$\ub{\psi}^o_{\beta, k}(x)$ and $\ub{\psi}^o_{\beta, k}(v)$
determined by \eqref{eq:v-b-trans}, \eqref{eq:wdw-x-def} fixes $B$
as
\begin{equation}\label{eq:wdw-norm-fact}
  B = \f{1}{\sqrt{2\pi k}} .
\end{equation}

To summarize, for every self-adjoint extension
$\ub{\Theta}'_{\Lambda,\beta}$ of $\ub{\Theta}'_{\Lambda}$ the
positive part of its spectrum $\Sp_+(\ub{\Theta}'_{\Lambda,\beta})$
equals the entire positive half $\re^+$ of the real line and is
absolutely continuous and non-degenerate. The domain $\Dom_{\beta}$
of the operator is spanned by eigenfunctions $\ub{e}_{\beta, k}$,
with $k\in\re^+$, given by \eqref{eq:wdw-ext-eigenf}, where the
coefficient $B$ is fixed by \eqref{eq:wdw-norm-fact}. The resulting
eigenfunctions satisfy the normalization condition
\begin{equation}
(\ub{e}_{\beta, k'}|\ub{e}_{\beta, k}) = \delta(k-k') .
\end{equation}

\subsection{The physical sector}
\label{s3.3}

Knowing the self-adjoint extensions of $\ub{\Theta}'_{\Lambda}$ one
can trivially construct all the self-adjoint extensions of $\ub{C}$
by substituting $\ub{\Theta}_{\Lambda}$ in \eqref{eq:C-wdw} with
$\ub{\Theta}_{\Lambda,\beta}=12\pi G\ub{\Theta}'_{\Lambda,\beta}$.
Then, knowing the spectral decomposition of
$\ub{\Theta}_{\Lambda,\beta}$ one can find the physical Hilbert
space corresponding to each extension by the group averaging
\cite{dm,almmt,abc}
method as in \cite{aps2}. The resulting (positive frequency)
physical states corresponding to each extension are of the form
\begin{equation}\label{eq:wdw-ph-state-def}
  \ub{\Psi}(x,\phi)
  = \int_{\re^+}\rd k\, \tilde{\ub{\Psi}}(k)\,
   \ub{e}_{\beta,k}(x)\, e^{i\omega(k)\phi} ,
\end{equation}
where $\omega(k)=\sqrt{12\pi G}\,k$. As explained in \cite{aps2,acs}
the physical inner product given by the group averaging procedure
reduces to simply
\begin{equation}
\langle \ub{\Psi}_1|\ub{\Psi}_2 \rangle = \int_{\re^+}\! {\rd} k\,
\ub{\bar{\tilde\Psi}}_1(k) \, \ub{\tilde\Psi}_2 (k) \quad {\hbox{\rm so that}}
\quad \Hil^{\phy} = L^2(\re^+,\rd k) .
\end{equation}
Knowing this form of physical states, we can introduce as usual
\cite{acs,as} the notion of the evolution with respect to the
internal time $\phi$, and interpret the Hamiltonian constraint as
the evolution equation. The resulting unitary evolution of the
initial data at $\phi=\phi_o$ is given by
\begin{equation}\label{eq:wdw-evo}
\ub{\Psi}(x,\phi_o) \mapsto
\ub{\Psi}(x,\phi) = e^{i(\phi-\phi_o)\,
\sqrt{|\ub{\Theta}_{\Lambda,\beta}|}}\,\,\ub{\Psi}(x,\phi_o)\, ,
\end{equation}
where $|\ub{\Theta}_{\Lambda,\beta}|$ is the positive part of
$\ub{\Theta}_{\Lambda,\beta}$.

To extract physics from this setup, one needs a family of relational
observables on $\ub{\Hil}_{\phy}$, parameterized by the internal
time $\phi$. In the cases $\Lambda\leq 0$ \cite{aps3,bp} this role
was served by operators $|\hat{v}|_{\phi}$ corresponding to the
volume (of the fiducial cell $\mathcal{C}$) at given value of
$\phi$. In our case, however, in the classical theory $v$ becomes
infinite at a finite value of $\phi$. On the quantum level this
property is reflected by the fact that an operator
$|\hat{v}|_{\phi}$ can map even the elements of the Schwartz space
$\Sch\in\Hil^{\phy}$ outside of $\Hil^{\phy}$. Since the physical
origin of this technical problem is clear from the classical
analysis, so is the solution: One can simply replace
$|\hat{v}|_{\phi}$ with $|f(\hat{v})|_{\phi}$, where $f$ is a
\emph{bounded} function of $v$. A convenient choice of $f$ is an
`angular parameter' $\theta \in ]0,\pi/2[$ defined by
\begin{equation}\label{eq:p-theta}
  V = K\, \tan(\theta) ,
\end{equation}
where $K$ is a constant with dimension of the volume. Following the
general procedure introduced in \cite{aps2,aps3}, one first
introduces a kinematical operator acting on the initial data space
$P^+_{\beta}\ub{\Hil}_{\gr}$ consisting of states in
$\ub{\Hil}_{\gr}$ in the positive part of the spectrum of
$\ub{\Theta}_{\Lambda, \beta}$
\begin{equation}\label{eq:wdw-theta-def}
\big(\hat{\theta}\ub{\psi}\big)(v) =
\big(P^+_{\beta}\,\arctan (\f{|v|}{K})\, P^+_{\beta}\, \ub{\psi}\big)(v)
\end{equation}
($P^+_{\beta}$ projects elements $\ub\psi (v)$ of $\Hil_{\gr}$ into
their positive frequency part by restricting the corresponding
Fourier transform $\ub{\tilde{\psi}}(k)$ to the positive half
$k$-line.) In the second step, one extends it to the physical
operator $\hat{\theta}|_{\phi}$ using \eqref{eq:wdw-evo}:
\be \big(\hat\theta|_{\phi_o}\, \Psi\big) (x,\phi) =
e^{i(\phi-\phi_o)\, \sqrt{|\ub{\Theta}_{\Lambda,\beta}|}}\,\,
\big(\hat\theta\ub{\Psi}\big) (x,\phi_o) \, , \ee
This relational Dirac observable enables us to effectively track the
evolution of volume of the fiducial cell $\mathcal{C}$.

There are other observables which  are manifestly independent of the
choice of a fiducial cell. An example is provided by the family
$|\hat{b}|_{\phi}$ built out of the kinematical observable
$|\hat{b}|$. This observable is of direct physical interest because
classically $b$ is proportional to the Hubble rate $H$:
$b=\gamma\lambda\, H$. Finally, another cell-independent observable
of direct physical interest is the total energy density
$\hat{\rho}_{\rm tot}|_{\phi}$ commonly used in LQC \cite{acs,klp1}.
It is constructed starting from the kinematical operator
\begin{equation}\label{eq:wdw-rho-def}
  \hat{\rho}_{\rm tot} = \frac{1}{2\lambda^2\,(2\pi\gamma G)^2}
  |\hat{v}|^{-1} \ub{\Theta}_{o} |\hat{v}|^{-1}
  = \frac{3}{8\pi G\,\lambda^2\gamma^2}\, \hat{b}^2 .
\end{equation}

We will now use this setup to explore quantum dynamics of the \WDW
theory.

\subsection{Quantum dynamics}
\label{sec:wdw-dynamics}\label{s3.4}

\begin{figure}[hbt!]
  \includegraphics[width=3.5in,height=2.4in]{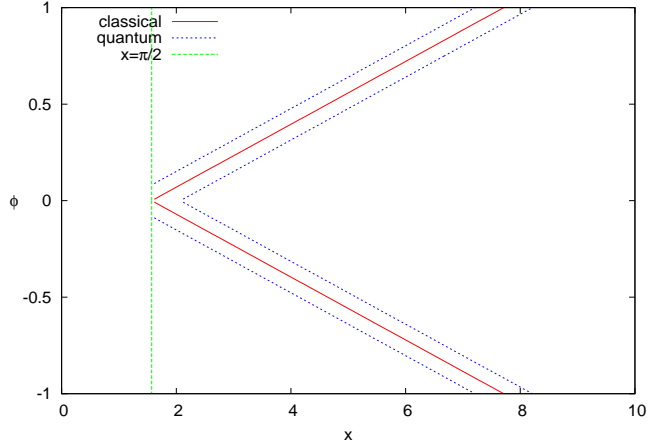}
 \caption{Qualitative behavior of the quantum evolution can be
inferred from the classical dynamics coupled with the form of
eigenfunctions. Expanding classical trajectories start at
$(x=\infty,\,\phi=-\infty)$ (the big bang) and end at
$(x=\pi/2,\, \phi=\phi_o)$ (where the matter density $\rho$ goes
to zero), while the contracting trajectories start at $(x=\pi/2,\,
\phi=\phi_o)$ and at $(x=\infty,\,\phi=\infty)$ (the big crunch).
For definiteness we have set $\phi_o =0$ in the figure. The form of
$\ub{\Theta}_{\Lambda,\beta}$ and its eigenfunctions in the
$x$-representation suggest that incoming semi-classical wave
functions would be largely reflected at $x=\pi/2$ and follow the
contracting classical trajectory which is the analytical continuation
of the original one. The figure shows only the part $x >0$ because
the physical states $\ub{\Psi}(x,\phi)$ are symmetric under reflection
in $x$.}
\label{fig:heuristic}
\end{figure}

Thanks to the simple form of the evolution operator
\eqref{eq:theta-wdw-x} and its eigenfunctions
\eqref{eq:wdw-ext-eigenf}, we can draw some general conclusions
regarding quantum dynamics at a semi-heuristic level. Note first
that, for $|x| > \pi/2$, every self-adjoint extension
$\Theta_{\Lambda, \beta}$ yields a standard Klein-Gordon equation in
the $x,\phi$ plane, with boundary conditions at the `barrier'
$|x|=\pi/2$ defined by that particular choice of $\beta$. Most of
the wave is reflected at this barrier as the tunneling amplitude is
exponentially suppressed. Therefore one would expect the qualitative
features of the evolution to be as in Fig.\ref{fig:heuristic}.
Consider a quantum state $\ub{\tilde\Psi}(k)$ which is sharply
peaked about a large $k^\star$ for which $\ub\Psi(x,\phi)$ is peaked
on a classical trajectory
\begin{equation}
  b(\phi) = b_o\,\cosh[\sqrt{12\pi G}(\phi-\phi_o)] \quad {\rm or}\quad
  x(\phi) = \f{\pi}{2} + \sqrt{12\pi G}\,(\phi_o-\phi)\,
\end{equation}
in the low curvature region at some $\phi <\phi_o$. Then, because of
the reflection at $|x| = \pm \pi/2$, one would expect
$\ub\Psi(x,\phi)$ to remain sharply peaked on the trajectory
\begin{equation}
  x(\phi) = \f{\pi}{2} + \sqrt{12\pi G}\,|\phi-\phi_o|\, .
\end{equation}
for all $\phi$, even to the future of $\phi=\phi_o$. But this is
only a qualitative argument because it neglects the tunneling into
the region $|x| < \pi/2$ and ignores the $k$ dependence of the
reflection coefficients.

To make precise statements regarding the global evolution one has to
analyze the expectation values and dispersions of appropriate
observables. Since the model is not analytically soluble, one has to
resort to numerical methods using a suitable class of states.
Following the existing literature, we select the Gaussians sharply
peaked about some large $k^\star$ and some $b^\star$ at given
initial time $\phi_o\ll -1/\sqrt{G}$. The spectral profiles
corresponding to such states are of the type
\be \label{eq:phistar} \ub{\tilde{\Psi}}(k) =
e^{-\f{(k-k^\star)^2}{2\sigma^2}}\, e^{-i\omega(k)\phi^\star} ,
\quad {\rm where} \quad \phi^\star := \phi_o + \f{1}{\sqrt{12\pi
G}}\, \arcosh(b^\star/b_o)\, .\ee
Since the explicit form of the basis functions $\ub{e}_{\beta,
k}$ is known via \eqref{eq:wdw-ext-eigenf} and
\eqref{eq:wdw-norm-fact} the wave function $\ub{\Psi}(b,\phi)$
can be calculated by first carrying out a direct integration of
\eqref{eq:wdw-ph-state-def} and then passing to the $b$
representation using \eqref{eq:wdw-x-def}).

\begin{figure*}[tbh!]
  \begin{center}
    $(a)$\hspace{3.2in}$(b)$
  \end{center}\vspace{-0.2cm}
  \includegraphics[width=3.2in]{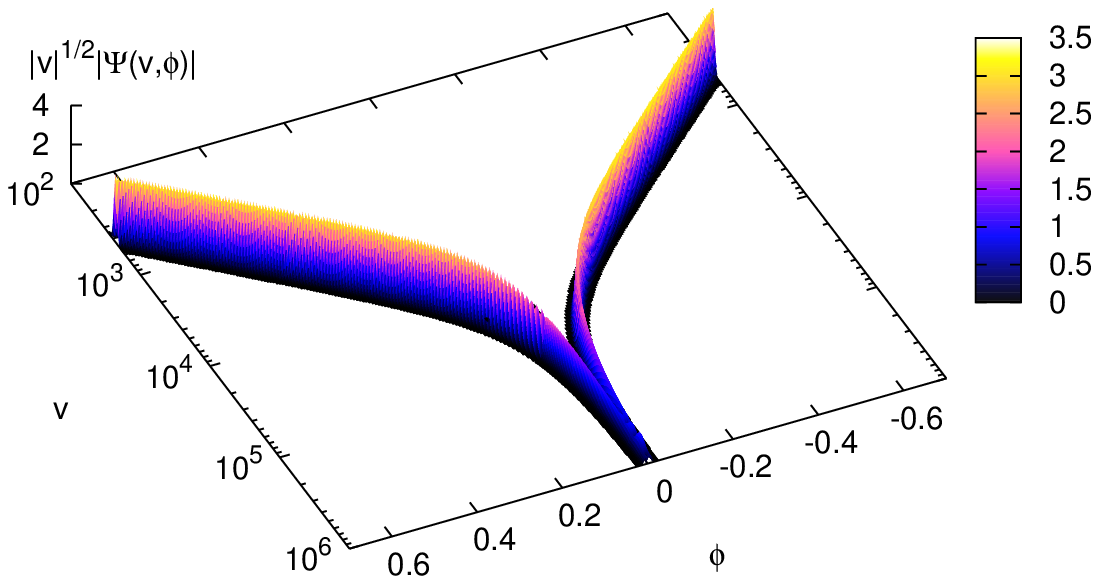}
  \includegraphics[width=3.2in]{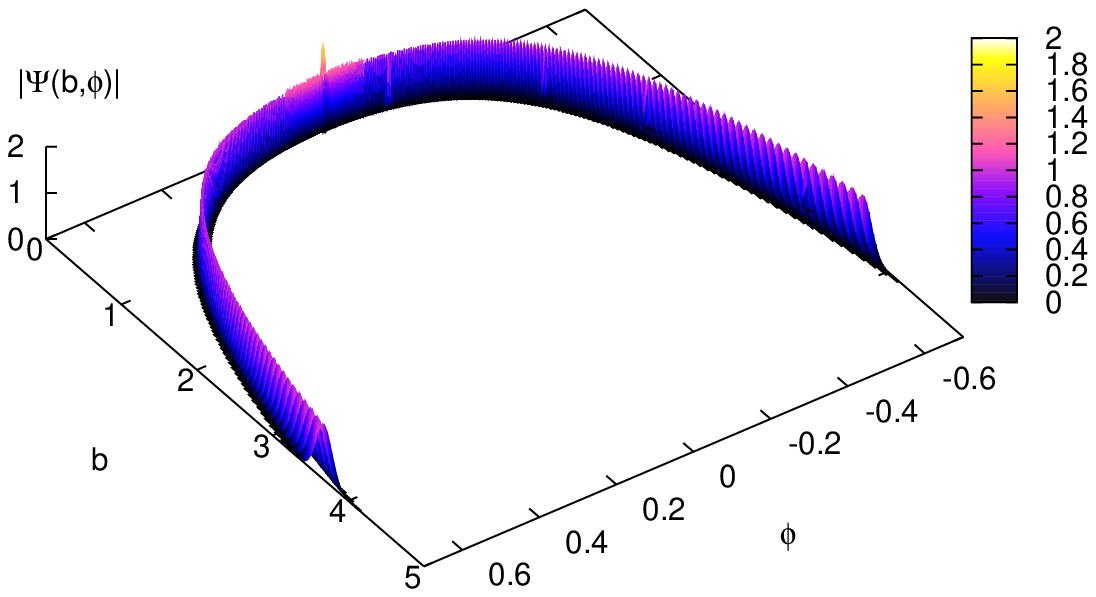}
  \begin{center}
    $(c)$\hspace{3.2in}$(d)$
  \end{center}\vspace{-0.2cm}
  \includegraphics[width=3.2in]{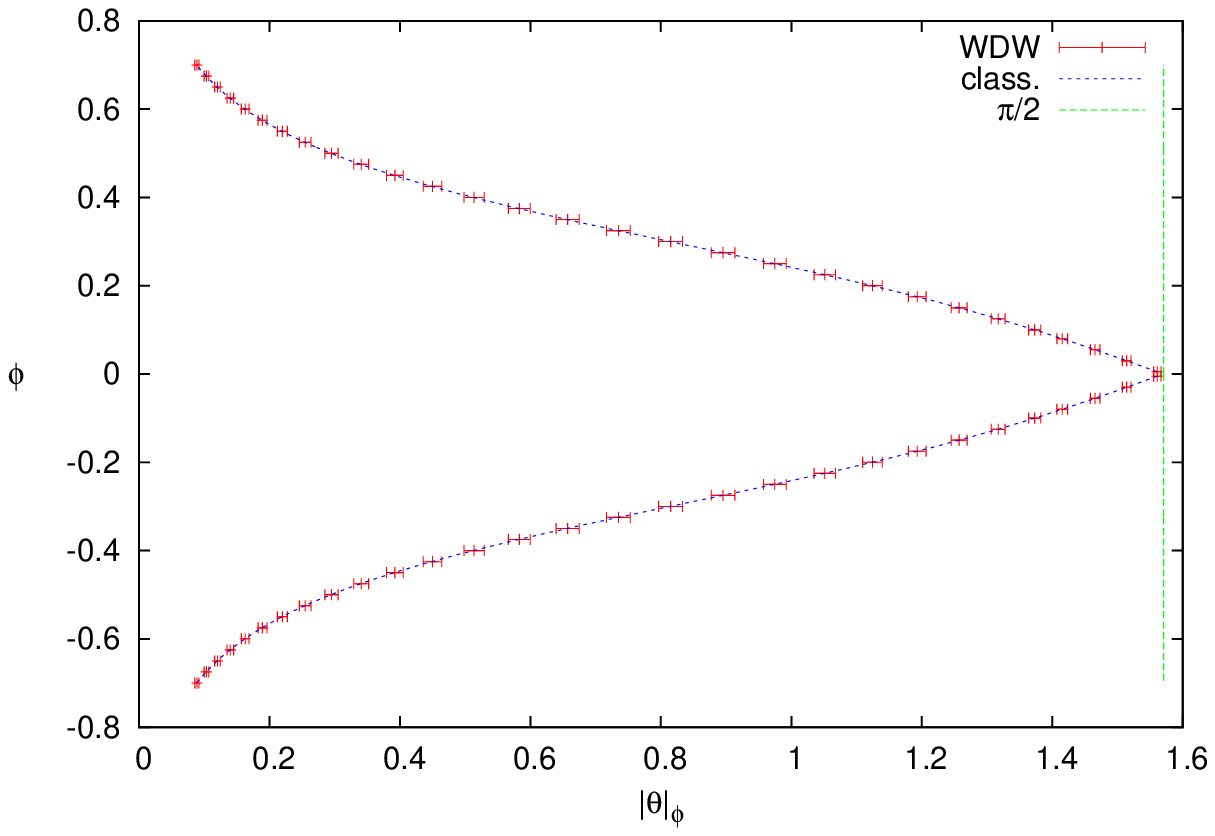}
  \includegraphics[width=3.2in]{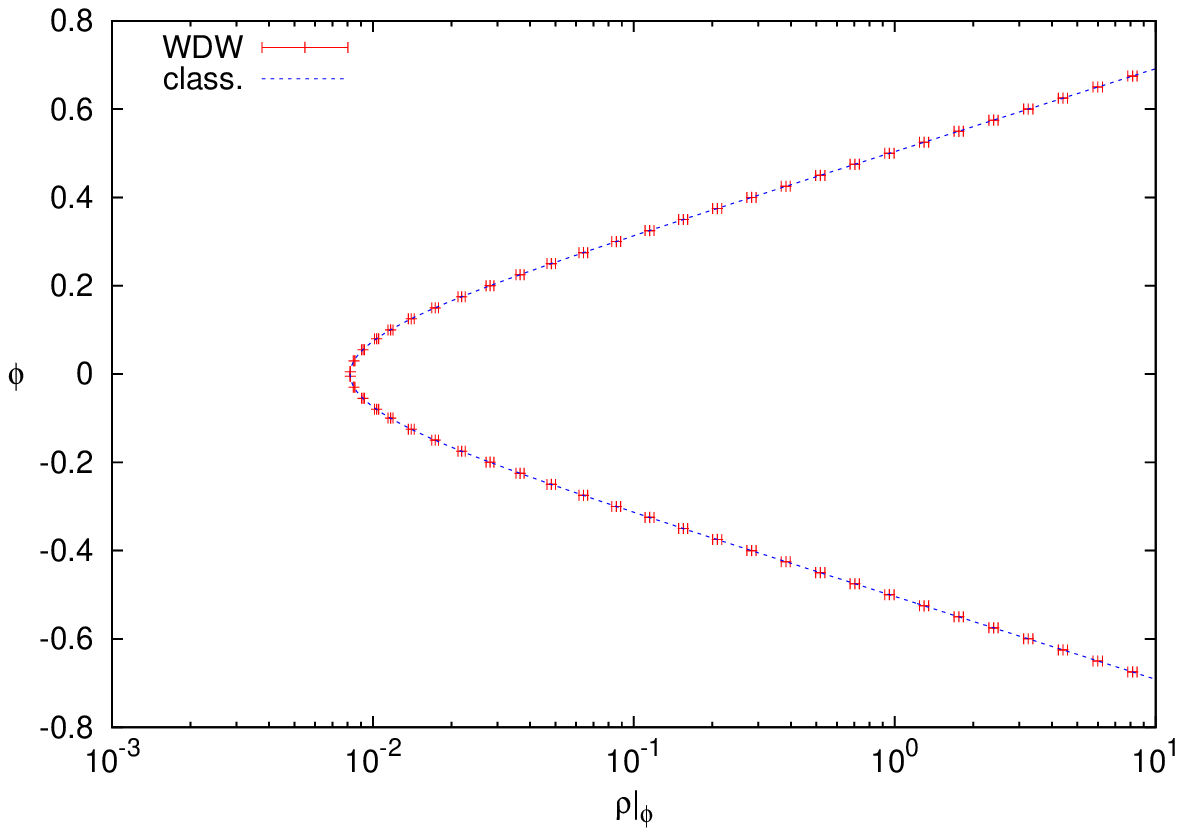}
  \caption{Quantum dynamics in the \WDW theory. Evolution of
    the wave packet initially peaked at
    $p_{\phi}^\star=5\times 10^3$ and $|\theta|^\star_{\phi=0}=\pi/2$ with
    dispersion $\Delta p_{\phi}/p_{\phi} = 0.03$. Figures (a) and (b)
    show the evolution in the $v$-$\phi$ and $b$-$\phi$ plane, respectively.
    Figures (c) and (d)  show the expectation values of the observables
    $|\hat{\theta}|_{\phi}$, and  $\hat{\rho}_{\rm tot}|_{\phi}$ in this state
    and compare them with the analytically extended classical trajectories.
    In these simulations, $\beta=0$, the constant $K$ in the
    definition of $\theta$ was set to $K= 5\times 10^3$, and $\Lambda = 0.2\lPl^{-2}$
  }
  \label{fig:WDW}
\end{figure*}

Recall that the expression of the inner product is quite complicated
in the $b$-representation but extremely simple in the
$v$-representation. On the other hand, the expression of some of the
basic operators is simpler in the $b$-representation. Therefore the
calculation requires us to pass between the two representations
using the Fourier transform \eqref{eq:v-b-trans} and its inverse.
Consequently, numerical analysis of dynamics ---calculation of
the expectation values and dispersions of $\hat{\theta}|_{\phi},
\hat{\rho}_{\rm tot}|_{\phi}$--- was carried out in the following
steps
\begin{enumerate}[(i)]
\item First, the profiles $\ub{\Psi}(x,\phi)$ at a given $\phi$
    were found via the integration of
    \eqref{eq:wdw-ph-state-def} over the interval $k \in
    [k^\star-7\sigma, k^\star+7\sigma]$ via
    Romberg method with the stepsize and order selected
    dynamically to achieve the relative integration precision of
    the order $10^{-6}$. The corresponding wave function
    $\ub{\Psi}(b,\phi)$ was then calculated using
    \eqref{eq:wdw-b-def}.
  \item Next, $\ub{\Psi}(b,\phi)$ was used to compute the function
    \begin{equation}
      \ub{\Phi}(b,\phi) := b^2 \ub{\Psi}(b,\phi) .
    \end{equation}
that is needed to evaluate the expectation values in step
(iv) below.
  \item Results were transformed to the $v$-representation via
      an inverse of \eqref{eq:v-b-trans}. The Fourier transform
      involved in it was computed via a Fast Fourier Transform
      (FFT) algorithm with $2^N$ probing points (where
      $N$ varied from $16$ to $19$). To adapt the resolution in
      $b$ (corresponding to the size of the domain of
      calculation in $v$) the domain of transformation
      $b\in[-b_M,b_M]$ was allowed to change in the process of
      evolution and varied from $0.5$ to $5$.
  \item Finally the norm, and the expectation values were
      calculated as follows
    \begin{subequations}\begin{align}
      \|\Psi\|^2 &= \int_{\re} \rd v |\Psi(v,\phi)|^2 , \\
      \obs{|\hat{\theta}|_{\phi}}
        &= \|\Psi\|^{-2} \int_{\re} \rd v\, \arctan(\f{|v|}{K}) |\Psi(v,\phi)|^2 , \\
      \obs{\hat{\rho}_{\rm tot}|_{\phi}} &= C_{\rho}
        \|\Psi\|^{-2} \int_{\re} \rd v\,
        \bar{\Psi}(v,\phi)\Phi(v,\phi) ,
    \end{align}\end{subequations}
    where 
    $C_{\rho}=3/(8\pi G\,\lambda^2\gamma^2)$. The
    dispersions were found via a standard relation
    \begin{equation}\label{eq:gen-disp-def}
      \delta_O^2 = \obs{\hat{O}^2} - \obs{\hat{O}}^2
    \end{equation}
 for a given observable $\hat{O}$. On the other hand, the
 expectation values of squared operators are determined via
    \begin{subequations}\begin{align}
      \obs{|\hat{\theta}|^2_{\phi}}
        &= \|\Psi\|^{-2} \int_{\re} \rd v\, \arctan^2(\f{|v|}{K}) |\Psi(v,\phi)|^2 , \\
      \obs{\hat{\rho}^2_{\phi}}
        &= C_{\rho}^2\|\Psi\|^{-2} \int_{\re} \rd v |\Phi(v,\phi)|^2 .
    \end{align}\end{subequations}
 All the integrations in this step were performed via the
 trapezoid method, with the configuration of the probing points
 preset by the FFT in the previous step.
\end{enumerate}

A large number of simulations were performed varying the parameters
of the initial state and the self-adjoint extension used in
evolution. They bore out the semi-heuristic expectation described in
the beginning of this sub-section. An example of the results is
presented in Fig.~\ref{fig:WDW}. For any choice of extension, states
that are sharply peaked in the distant past (or, future) remain so
throughout the evolution and follow the analytically extended
classical trajectories towards big bang and big crunch
singularities. \emph{Thus, as in the $\Lambda =0$ case, the big bang
and the big crunch singularities are not resolved in the \WDW
theory.} The new element is that the wave packets follow the
\emph{analytically extended} classical trajectory beyond
$\phi=\phi_o$, \emph{irrespective of the choice of the self-adjoint
extension.}%
\footnote{A previous result in the \WDW theory, but using
  a `spinor formalism' \cite{k-poLwdw}, hinted at the existence
  of such an extension. Although the scale factor played the role
  of time there, the quantum wave packet developed a second
  branch which, in retrospect, may be interpreted as the analog
  of this extension.
}
%

Now, usually in quantum mechanics the existence of inequivalent
self-adjoint extensions implies that at some point the physical
evolution breaks down and additional data are required to
continue it. In our case these data correspond to the gluing
conditions at $|b|=b_o$ or $|x| = \pi/2$. However,
surprisingly, for states which start out as semi-classical in
the weak curvature regime, this non-uniqueness does not
manifest itself at least at the level of the expectation values
of the observables: All of them follow the (analytically
extended) classical trajectories dictated by parameters
$k^\star, b^\star, \phi^\star$ used in their construction. This
result is a reflection at the quantum level of the fact that
the classical solutions admit a \emph{unique} analytic
extension.

But what about generic wave functions? To better understand the
dependence of physical quantities on the choice of extension in this
case, let us first consider wave functions in the $x$ variable. For
$|x|\geq\pi/2$ each basis function $\ub{e}_{\beta, k}$ has the form
of a standing wave and can be decomposed onto the incoming
$\ub{e}^-_{\beta k}$ and outgoing $\ub{e}^+_{\beta, k}$ component
\begin{equation}
  \ub{e}^{\pm}_{\beta k} = (1/\sqrt{8\pi k})\,
  e^{\pm i(k|x|+\sigma_{\beta}(k))} .
\end{equation}
By replacing $\ub{e}_{\beta, k}$ in \eqref{eq:wdw-ph-state-def} with
$\ub{e}^{\pm}_{\beta, k}$, we arrive at a split of the wave function
$\ub{\Psi}(x,\phi)$ onto expanding $\ub{\Psi}^-$ and contracting
$\ub{\Psi}^+$ components. Let us define
\be \ub{e}'^{\pm}_{k}(x) = e^{\mp i\sigma_{\beta}(k)}\,\,
\ub{e}^{\pm}_{\beta, k}(x) \ee
by rotating each basis function so that the result
$\ub{e}'^{\pm}_{k}(x)$ is $\beta$-independent and rewrite the
expanding and contracting wave functions in terms of them:
\be \ub{\Psi'}^{\pm}(x,\phi) = \int_{\re^+}\rd k\,
\tilde{\ub{\Psi}}'^{\pm}(k) \, \, \ub{e}'^{\pm}_{k}(x) \,
e^{i\omega(k)(\phi-\phi_o)}\, .\ee

Then the spectral profiles $\tilde{\ub{\Psi}}'^{\pm}$ satisfy
\be \tilde{\ub{\Psi}}'^{\pm}(k) = e^{\pm i\sigma_{\beta}(k)}
\tilde{\ub{\Psi}}(k)\, ,\quad {\rm whence} \quad
\tilde{\ub{\Psi}}'^{+}(k) = e^{2i\sigma_{\beta}(k)}
\tilde{\ub{\Psi}}'^{-}(k)\, . \ee
Thus upon reflection, the initial expanding wave profile
$\ub{\Psi}^{'-}(k)$ undergoes a phase shift
$e^{2i\sigma_{\beta}(k)}$ that depends on the choice of the
self-adjoint extension. Note that the expression
\eqref{eq:wdw-ext-rot} of $\sigma_{\beta}$ implies that the
difference $\sigma_{\beta}-\sigma_{\beta'}$ in rotations of the
phase caused by two different extensions is not global, but
depends on $k$. This subtlety should be reflected in the
evolution of observables.

To extract this information, it is convenient to follow
\cite{kp-scatter} and  regard the global evolution as a scattering
process. For, in the distant past, i.e., $\phi \ll - 1/\sqrt{G}$,
and in the distant future, i.e. $\phi \gg 1/\sqrt{G}$, we are near
the big bang and big crunch singularities where the effect of the
cosmological constant can be neglected relative to that of matter,
i.e., $\Lambda/8\pi G \ll \rho$. Therefore, in these `asymptotic'
regions, dynamics is better and better approximated by that in the
$\Lambda =0$ case. Thus we can regard `incoming' and `outgoing
states' as belonging to the $\Lambda=0$ physical Hilbert spaces. The
incoming state can be thought of as being scattered because of the
presence of the cosmological constant which dominates dynamics near
$|x| =\pi/2$. As in \cite{kp-scatter}, the scattering process can be
analyzed using observables $\ln|v|_{\phi}$. However, it is more
convenient to consider observables $\ln|\hat{b}|_{\phi}$ defined in
analogy to $|\hat{b}|_{\phi}$. The limit of the basis functions is
provided already by \eqref{eq:wdw-e-limit}. Following our analysis
in the $x$ variable, we split this limit into incoming and outgoing
components expressed in terms of the basis of $\ub{\Theta}_o$ in the
$b$-representation
\begin{subequations}\begin{align}
    \ub{e}_{\beta, k}(b) &=\f{1}{\sqrt{2}}\,\,
    e^{i(k\beta_o+\sigma_{\beta}(k))}\, {\ub{e}}^+_{k}(b)\, 
+ e^{-i(k\beta_o+\sigma_{\beta}(k))}\, \ub{e}^-_{-k}(b)\, +\,
O(b^{-2})\, , \\
 {\rm with} \quad \ub{e}_{k}(b) &= (1/\sqrt{4\pi|k|}) e^{ik\ln|b|} ,
\end{align}\end{subequations}
where $\beta_o := \pi/2+\ln(2/b_o)$. Using this split in
\eqref{eq:wdw-ph-state-def} we can define the asymptotic future and
past states $\ub{\Psi}^{\pm}$ of the spectral profiles:
\begin{equation}
  \tilde{\ub{\Psi}}^{\pm}(k)
  = \theta(\mp k)\, e^{\pm i(|k|\beta_o+\sigma_{\beta}(|k|))}
\, \tilde{\ub{\Psi}}(|k|)\,  .
\end{equation}
Now the global evolution is described by the scattering operator
$\h{S}$
\begin{equation}\label{eq:wdw-scatter}
  \ket{\ub{\Psi}^-}\,\,\, \mapsto\,\,\, \ket{\ub{\Psi}^+} =
\hat{S} \ket{\ub{\Psi}^-}
\end{equation}
where the matrix elements of the scattering operator $\hat{S}$ are
given by
\be S(k,k') := (\ub{e}_{k}|\,\hat{S}\,|\ub{e}_{k'})
  \,=\, e^{-2i\sgn(k')(\beta_o+\sigma_{\beta}(|k|))}\,\, \delta(k+k') .
\ee
The extension dependence is encoded entirely in the phases shift
$\sigma_{\beta}(k)$ given by \eqref{eq:wdw-ext-rot}. For large $k$,
it can be expanded as
\begin{equation}\label{eq:wdw-sigma-exp}
  \sigma_{\beta}(k) = -\frac{k\pi}{2} + \frac{\pi}{2} - \beta
  - \frac{1}{2}\sin(2\beta) e^{-k\pi} + O(e^{-2k\pi}) \, .
\end{equation}

Let us now select two extensions corresponding to some $\beta$ and
$\beta'$ and two states $\ket{\ub{\Psi}}, \ket{\ub{\Psi}'}$
corresponding to those extensions, sharply peaked at some $k^\star$
and such that in the asymptotic past they are equal. Then by
\eqref{eq:wdw-scatter} in the asymptotic future they are related
just by a phase rotation
\begin{equation}
  \tilde{\ub{\Psi}}^+(k) = e^{-2i(\sigma_{\beta}(k)-
  \sigma_{\beta'}(k))}\,\, \tilde{\ub{\Psi}}'^+(k) .
\end{equation}
This allows us to estimate via \eqref{eq:wdw-sigma-exp} the leading
order correction to the difference between the expectation values of
$\ln|\hat{b}|_{\phi}$ as
\be \label{compare}
|\bra{\ub{\Psi}^+}\,\ln|\hat{b}|_{\phi}\,\ket{\ub{\Psi}^+} -
\bra{\ub{\Psi}'^+}\,\ln|\hat{b}|_{\phi}\,\ket{\ub{\Psi}'^+}|  =
\pi\cos(\beta+\beta')\sin(\beta-\beta')\, e^{-\pi k^{\star}} +\,
O(e^{-2\pi k^{\star}})\,  . \ee
(Since $\ub{\Psi^+}$ and $\ub{\Psi}^{'+}$ are asymptotic states,
i.e. states in the $\Lambda=0$ theory, the difference is
$\phi$-independent.) To appreciate the physical implication of
this result, let us focus for a moment on semi-classical states
peaked at a large $k^\star$ such that initially the dispersions in
$\ln|b|$ and $\ln|k|$ are comparable and the uncertainty product is
approximately saturated. Then, in either the $\beta$ or $\beta'$
theory, because of the Heisenberg uncertainty principle, the
dispersion in $\ln |\hat{b}|$ is of the order of
$|k^{\star}|^{-(1/2)}$. Because of the exponential suppression in
(\ref{compare}), the difference in the expectation values of $\ln
|\hat{b}|$ in the $\beta$ and $\beta'$ theories is therefore
completely negligible compared to the dispersion in $\ln |\hat{b}|$
in either of these theories.

Let us now return to general states. The dependence of dispersions
of the choice of self-adjoint extension is slightly stronger than
that in (\ref{compare}). Results of \cite{kp-scatter} show that the
possible growth of dispersion between the asymptotic past and future
states depends on the behavior of $\partial_k \sigma_{\beta}(k)$. By
repeating for the \WDW theory the derivation of certain triangle
inequalities on dispersions obtained in \cite{kp-scatter} for LQC,
one can shows that the expected difference between the dispersions
$\delta_+$ and $\delta_+'$ of $\ln|b|$ for the two states
$\ket{\ub{\Psi}^+}$, $\ket{\ub{\Psi}'^+}$ (of the \WDW theory now
under consideration) will be of the order of
\begin{equation}\begin{split}
|\delta_+-\delta_+'| &\, \lessapprox\,
\bra{\ub{\Psi}^+}\,\Delta(\partial_k
(\sigma_{\beta}-\sigma_{\beta'}))\,\ket{\ub{\Psi}^+} \\
& \,\lessapprox |\,[\partial^2_k (\sigma_{\beta}-\sigma_{\beta'})]\,(k^\star)|
\,\sigma_{\ln|k|} \\
&\, \lessapprox \pi^2\,|\,\cos(\beta+\beta')\sin(\beta-\beta')\,|\,\,
e^{-\pi k^{\star}}\,\delta_{\ln|k|} .
\end{split}\end{equation}
where $\delta_{\ln|k|}$ is the dispersion of the observable
$\ln|\hat{k}|$. (Here and in what follows the approximate sign
in the inequality emphasizes the fact that we are keeping track
only the leading order terms.)

Now, the change in the dispersion between the asymptotic past and
future state is given by
\begin{equation}\begin{split}
|\delta_+-\delta_-| &\lessapprox \,
2 \bra{\ub{\Psi}^+}\,\Delta(\partial_k \sigma_{\beta})\,
\ket{\ub{\Psi}^+} \\
&\, \lessapprox 2|\,[\partial^2_k \sigma_{\beta}](k^\star)\,|\,\,
 \delta_{\ln|k|} \\
&\lessapprox \pi^2\sin(2\beta)\, e^{-\pi k^{\star}}\,\delta_{\ln|k|} ,
\end{split}\end{equation}
where $\delta_-$ is the dispersion of the observables
$\ln|\hat{b}|_{\phi}$ on the state $\ket{\ub{\Psi}^-}$. (Again since
the asymptotic states refer to the $\Lambda=0$ theory, this
difference is $\phi$ independent.) Thus, the difference in the
dispersions $|\delta_+-\delta_+'|$ in the asymptotic future in the
$\beta$ and $\beta'$ theory is comparable to the change in the
dispersion $|\delta_+-\delta_-|$ between asymptotic past and future
in any one theory. Both these quantities are \emph{very} small
compared to the dispersion $\delta_{\ln |k|}$ in the scalar field
momentum.

To summarize, computer simulations of states which start out as
Gaussian in the weak curvature region established that the big bang
and the big crunch singularities fail to be resolved in the \WDW
theory. They also brought out the surprising fact that although the
final physical sector of the quantum theory does depend on the
choice of the self-adjoint extension $\ub{\Theta}_{\Lambda,\beta}$
of the symmetric operator $\ub{\Theta}_{\Lambda}$, the difference in
the dynamics of these states is negligible. The S-matrix strategy
first introduced in \cite{kp-scatter} enabled us to obtain certain
analytical inequalities for generic states for which the dispersion
in $k$ ---i.e., in the field momentum--- is finite. They showed that
if the dispersion $\delta_{\ln |k|}$ is small, then differences in
the S-matrix predictions of physical theories that result from
different choices of self-adjoint extensions are enormously
suppressed.

\section{Loop Quantum Cosmology}
\label{s4}

In this section we will show that the procedure followed in Sec.
\ref{s3}  for the \WDW theory can be repeated in a rather
straightforward manner for LQC. Therefore our discussion will be
parallel to that of Sec.~\ref{sec:wdw-rep} and
\ref{sec:wdw-dynamics} with emphasis on the differences from the
\WDW theory.

\subsection{LQC kinematics}\label{sec:lq}\label{sec:lq-kin}

As in the \WDW theory, the kinematical Hilbert space is a tensor
product $\Hilk = \Hil_{\gr}\otimes \Hil_{\phi}$. However, while we
again have $\Hil_{\phi} = L^2(\re, \rd \phi)$, the gravitational
Hilbert space is now different \cite{abl,as}. As in the $\Lambda=0$
case \cite{aps3}, it is given by $\Hil_{\gr} =
L^2(\bar{\re},\rd\mu_{\Bohr})$, where $\bar{\re}$ is a Bohr
compactification of the real line and $\rd\mu_{\Bohr}$ the Haar
measure thereon. A convenient basis is again provided by the
eigenvectors of the operator $\hat{v}$:
\begin{equation}\label{eq:v-def}
  \hat{v}\ket{v} = v\ket{v}, \quad{\hbox{{\rm so that}}}\quad
  \hat{V}\ket{v}= (2\pi\gamma\lambda\lPl^2)\, |v|\, \ket{v}\, ,
\end{equation}
where, as before, $\gamma$ is the Barbero-Immirzi parameter of LQG
and $\lambda^2$ is the LQC \emph{area gap} \cite{awe1}. As in the
\WDW theory, in the volume representation states in $\Hil_{\gr}$
become wave functions $\psi(v)$, which are again taken to be
symmetric $\psi(v) = \psi(-v)$ to incorporate the fact that $v\to
-v$ is a large gauge transformation corresponding to the flip of the
orientation of the physical triad. However, unlike in the \WDW
theory, the $\psi(v)$ now have support only on a countable set of
points along the $v$-axis and their inner product is given by a sum
\begin{equation}\label{eq:ip-gr}
  \langle\psi|\psi'\rangle = \sum_{v\in\re}\,\, \bar{\psi}(v)
  \psi'(v)\, ,
\end{equation}
rather than an integral.

Because the Hilbert space is so different, the differential operator
$\ub{\Theta}_o = -12\pi G\, \sqrt{|v|}\,
\partial_v\, |v|\, \partial_v\sqrt{|v|}$ of the \WDW theory
fails to be well-defined on $\Hil_{\gr}$. Therefore, one has to
first express the classical constraint in terms of the
elementary variables of LQG ---holonomies and fluxes--- and
then promote the result to an operator on $\Hil_{\gr}$. This
systematic procedure leads to the following form of the
constraint operator \cite{aps3}
\begin{equation}\label{eq:constr-quant}
  \hat{C} = \id\otimes\partial_{\phi}^2 + \Theta_{\Lambda} \otimes \id ,
  \qquad \Theta_{\Lambda} := \Theta_o - \pi G\gamma^2\lambda^2\,\Lambda\,
  v^2\, ,
\end{equation}
where
\begin{equation}\label{eq:ev2-gen}
-[\Theta_o\psi](v) = f_+(v)\,\psi(v-4) - f_o(v)\,\psi(v)+ f_-(v)\psi(v+4)\, ,
\end{equation}
with the coefficients $f_{o,\pm}$ given by
\be \label{eq:fslqc} f_\pm(v) = (3\pi G/4)\, \sqrt{v(v\pm 4)}\,
(v\pm 2), \qquad
  f_o(v) = (3\pi G/2) v^2 \, .
\ee
Thus, the second order differential operator $\ub{\Theta}_o$ of
the \WDW theory is now replaced by a second order difference
operator $\Theta_o$ with uniform steps of size $v=\pm 4$.
Therefore, there is super-selection: one can investigate
dynamics separately on uniform lattices in the $v$-space and
each sector consisting of wave functions with support on any
one of these lattices is preserved by the complete set of Dirac
observables of interest, discussed in section \ref{s3}. In this
paper, we will restrict ourselves to the lattice $\lat =
\{v=4n,\, n\in\integ\}$ for simplicity because in LQC physical
results are largely insensitive to the choice of the sector
\cite{mop}.

Finally, as in the \WDW theory, for technical reasons it is
more convenient to work in the dual representation in which
states are wave functions $\psi(b)$ of the conjugate variable
$b$. However, since the LQC states $\psi(v)$ have support only
at $v= 4n$, we now have a Fourier series in place of the
Fourier integral \eqref{eq:v-b-trans}:
\begin{equation}\label{eq:lqc-v-b-trans}
  [\Fou\psi](b) =
  \frac{1}{2\sqrt{\pi}}\, \sum_{\lat_0\setminus\{0\}} |v|^{-\f{1}{2}}\,
  \psi(v)\, e^{\f{i}{2}vb} ,
\end{equation}
where the point $v=0$ was removed from the transform because
the state with support just at $v=0$ is dynamically decoupled
from the orthogonal sub-space spanned by states which vanish at
$v=0$. Since $\psi$ are supported on $\lat_0$, their images
$\Fou\psi$ are periodic in $b$ with the period $\pi$. Therefore
one can restrict the support of the wave functions
$[\Fou\psi](b)$ just to the circle $b\in[0,\pi]$, with the
identification $[\Fou\psi](0)=[\Fou\psi](\pi)$.

By inspection, the elementary operators $\hat{v}$ and
$\hat{\mathcal{N}}_{\mu}$ defined by
\begin{equation} \label{ops}
  \hat{v}\ket{v} = v\ket{v} , \qquad {\rm and} \qquad
  \hat{\mathcal{N}}_{\mu}\ket{v} = \ket{v+\mu} ,
\end{equation}
in the $v$ representation are transformed to
\begin{equation}
  \hat{v} = 2i\partial_b , \qquad {\rm and} \qquad
  \mathcal{N}_{\mu} = e^{-i\mu b/2} .
\end{equation}
in the $b$ representation. As a consequence, the operator
$\Theta_{\Lambda}$ assumes the form
\begin{equation}
  \Theta_{\Lambda} = - 12\pi G\,\, \big[\, (\sin(b)\partial_b)^2 -
  b_o^2\partial_b^2 \big],
\end{equation}
in the $b$ representation, where, as before $b_o:=
\gamma\lambda\, \sqrt{\Lambda/3}$.

Let us first consider the case when $b_o \ge 1$ or $\Lambda \ge
\Lambda_c:= 3/\gamma^2\lambda^2$. In this case,
$\Theta_\Lambda$ is essentially self-adjoint, whence one can
readily repeat the procedure of section \ref{s3.3} to construct
the physical Hilbert space. However, because this
$\Theta_\Lambda$ is negative, the physical Hilbert space is now
zero dimensional! (For proofs, see \cite{kp-posL}.) Thus, in
striking contrast to the \WDW theory, in LQC a non-trivial
quantum theory exists \emph{only when the cosmological constant
$\Lambda$ is less than a critical value, $\Lambda_c$.} Although
this result is not phenomenologically relevant because
$\Lambda_c$ is of Planck scale, it is of considerable
conceptual interest. In the rest of this section, then, we will
with work $\Lambda < \Lambda_c$.

\subsection{Properties of $\Theta_{\Lambda}$}
\label{sec:LQC-Theta-prop}

\subsubsection{Weak solutions to the eigenvalue equation}

\begin{figure*}[tbh!]
  \begin{center}
     $(a)$\hspace{3.2in}$(b)$
  \end{center}
  \includegraphics[width=3.2in,height=2.4in]{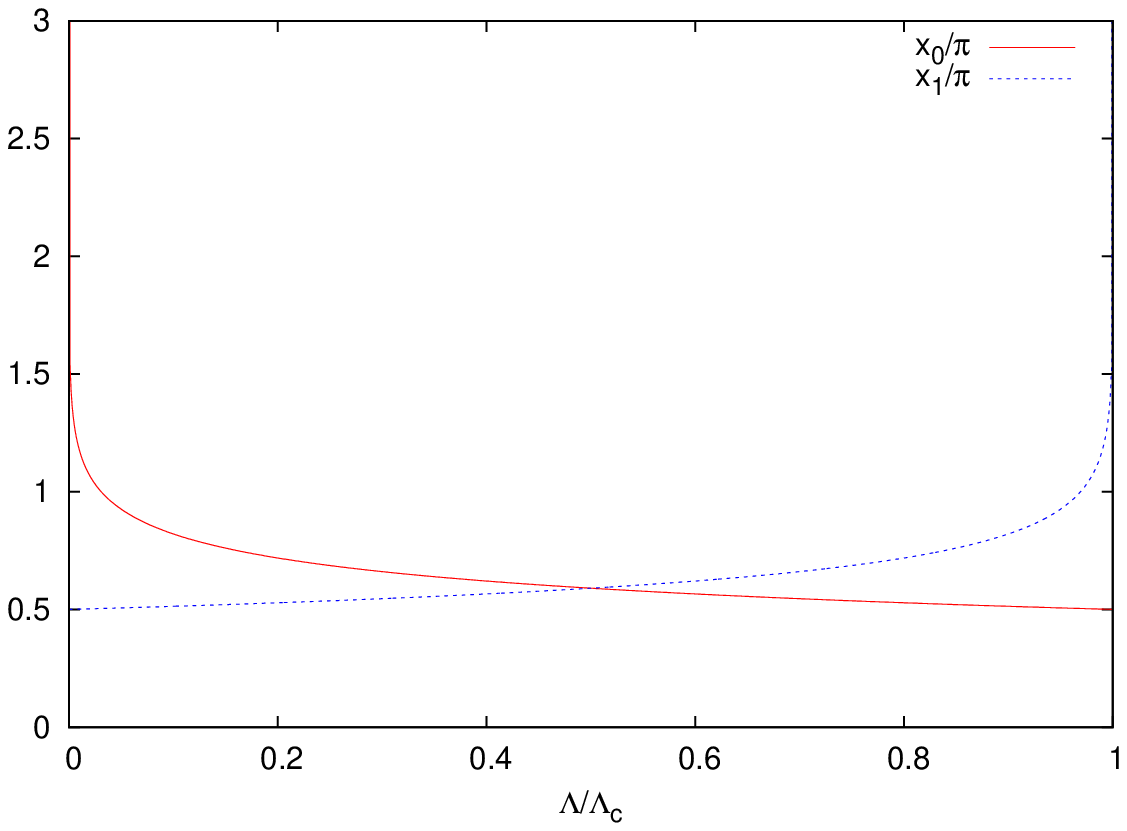}
  \includegraphics[width=3.2in,height=2.4in]{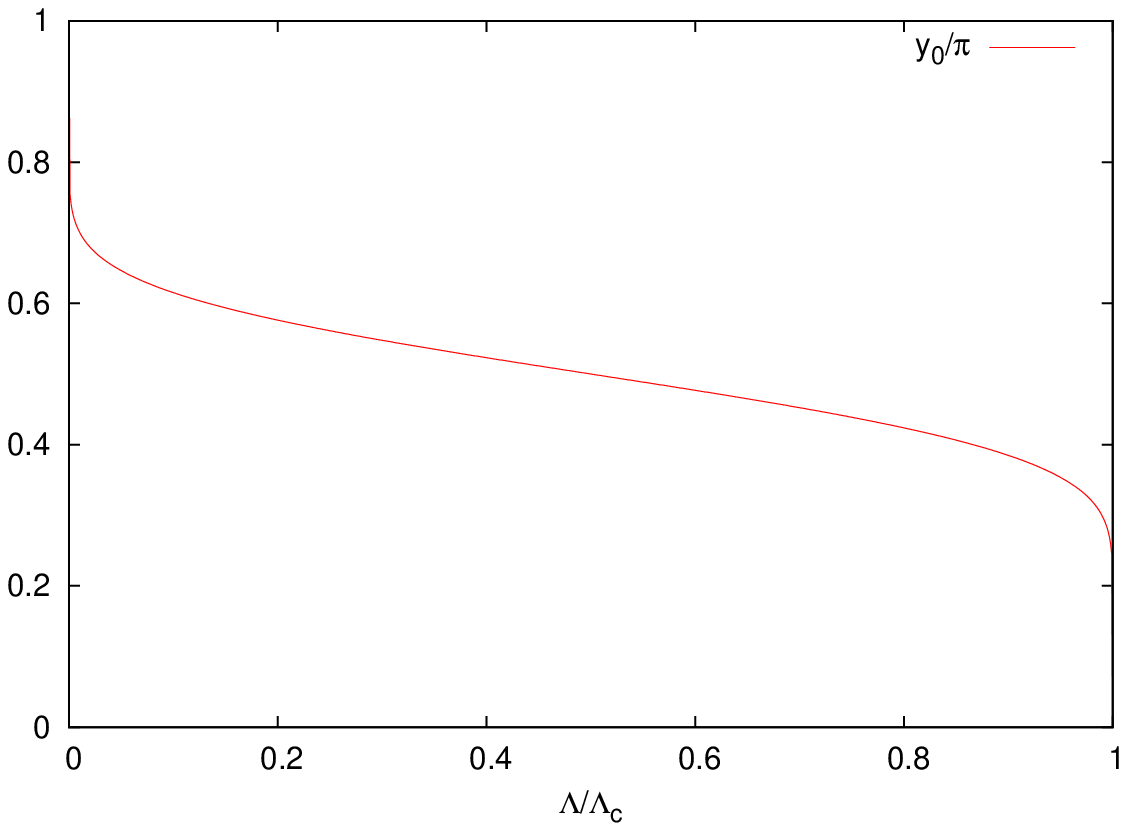}
  \caption{$(a)$\, The dependence of the functions $x_o$ and
  $x_1:=x_M-x_o$ (of \eqref{eq:xoM-def}) on the cosmological
  constant $\Lambda$. $(b)$ The dependence of $y_o = \pi x_o/x_M$
  on $\Lambda$. Here $\Lambda_c = 3/\gamma^2\lambda^2$ is the
  critical value of $\Lambda$ at which the energy density in the
  cosmological constant equals the maximum energy density
  $\rcr$ allowed by LQC in this model.}
  \label{fig:xoM}
\end{figure*}

Note that, in the $b$-representation, the gravitational part of the
Hamiltonian constraint is a differential operator as in the \WDW
theory. This suggests that one may be able to simplify it by a
change of variables. We will now show that $\Theta_\Lambda$ can in
fact be transformed to the \emph{same} form as in the \WDW theory:
\begin{equation}\label{eq:lqc-theta-x}
\Theta_{\Lambda} = 12\pi G\, \sgn(|x|-x_o)\,\,\partial_x^2 \, .
\end{equation}
However, there are two key differences. First, now the new variable
takes values on a compact interval $x\in[-x_M,x_M]$, with points
$-x_M$ and $x_M$ identified. Second, the transformation is much more
complicated in that $x$ is defined in terms of the elliptic integral
of the first kind $F(y,k)$:
\begin{equation}\label{eq:lqc-x-def}
  x:= \begin{cases}
\frac{1}{\sqrt{1-b_o^2}}\,\, F(b',\,1/(1-b_o^2)) ,& |b'|< \b_o , \\
x_M -\frac{1}{b_o}\, F(\pi/2-b',1/b_o^2) , & b' > \b_o ,\\
-x_M + \frac{1}{b_o}\, F(\pi/2+b',1/c^2) , & b' < -\b_o ,
      \end{cases}
\end{equation}
where
\ba \label{eq:xoM-def}
 b' := b - \pi/2, &\qquad& \b_o := \arcsin(b_o)\nonumber\\
x_o := \frac{1}{\sqrt{1-b_o^2}} F(\pi/2-\b_o,\,1/(1-b_o^2)),
&\,\,&
  x_M := x_o + \frac{1}{b_o} F(\b_o,\, 1/b_o^2)\, . \ea
The dependence of $x_o$ and $x_M$ on $b_o$ is shown in
Fig.~\ref{fig:xoM}. In particular
\begin{subequations}\begin{align}
  \lim_{b_o\to 0} x_o(b_o) &= +\infty , &  \lim_{b_o\to 0}
  [x_M-x_o](b_o) &= \pi/2 , \\
  \lim_{b_o\to 1} x_o(b_o) &= \pi/2 , & \lim_{b_o\to 1}
  [x_M-x_o](b_o) &= +\infty .
\end{align}\end{subequations}

For the technical simplicity, it is convenient to introduce the
rescaled variable
\begin{equation}
   y := \pi\, \f{x}{x_M}\,\,\, \in [-\pi,\pi],
\end{equation}
in terms of which the operator $\Theta_{\Lambda}$ takes the
form
\begin{equation}
  \Theta_{\Lambda} = \frac{12\pi^3 G}{x_M^2}\,\, \Theta'_{\Lambda} , \quad
{\rm where} \quad   \Theta'_{\Lambda} := \sgn(|y|-y_o)\, \partial_y^2 ,
\end{equation}
with $y_o := \pi x_o/x_M$.

We are now ready to analyze weak eigenfunction $\psi_{{\zeta}}$ of
$\Theta'_{\Lambda}$. As in the \WDW analysis we will have to go
back and forth between the $v$ and the $y$ representations. The
weak eigenfunctions are distributional solutions to
\begin{equation}\label{eq:lqc-eigenf}
    (\psi_{{\zeta}}|\Theta'^\dagger\, -\,\bar{{\zeta}}
  \id\,\ket{\chi} = 0 \qquad \forall \chi\in\Dom\ \, ,
\end{equation}
where $\Dom$ now consists of states $\chi \in \Hil_{\gr}$ which have
support only on a finite number of points of the lattice $\lat$ on
the v-axis. The particular form of the coefficients $f_{o,\pm}$ of
$\Theta_o$ in Eq. \eqref{eq:fslqc} and the form of the inner product
of $\Hil_{\gr}$ in the $v$-representation allows one to split each
such eigenfunction $\psi_{{\zeta}}$ into components $\psi_{{\zeta}}^{\pm}$
\begin{equation}
  \psi_{{\zeta}}^{\pm}(v) := \theta(\pm v)\, \psi_{{\zeta}}(v) ,
\end{equation}
which again satisfy \eqref{eq:lqc-eigenf} as in the \WDW theory.
Secondly, as a function of $y$, any solution to
\eqref{eq:lqc-eigenf} is necessarily of the form
\begin{equation}\label{eq:lqc-eig-sol}
  {\psi}_{{\zeta}}(y)\,=\,\,
   {\begin{cases}
    A (e^{i\sqrt{\zeta}y}  +  e^{-i\sqrt{\zeta}y}), & |y| \in [0,y_o) \\
    B (e^{\sqrt{\zeta}(\pi-y)}  +  e^{-\sqrt{\zeta}(\pi-y)}), & |y| \in (y_o,\pi] \\
  \end{cases}}
\end{equation}
analogous to \eqref{eq:eig-sol}. These two observations allow us to
directly apply the techniques developed in the final part of
Sec.~\ref{sec:wdw-rep} to find the necessary and sufficient
conditions for \eqref{eq:lqc-eig-sol} to satisfy
\eqref{eq:lqc-eigenf}. The result is a complete analog of the one in
the \WDW theory: the weak eigenfunctions have to be continuous in
$y=\pm y_o$ but not necessarily differentiable there. This property
transfers directly to the operator $\Theta_{\Lambda}$ for which any
(weak) eigenfunction is again globally continuous, but not
necessarily differentiable with respect to $x$ at $x=\pm x_o$.

\subsubsection{The self-adjoint extensions of $\Theta_\Lambda$}

We know already from the analysis of \cite{kp-posL} that
$\Theta_\Lambda$ admits a one parameter family of self-adjoint
extensions, labeled by elements of $U(1)$. Each extension
corresponds to particular asymptotic behavior of the basis functions
$e(v)$ in the limit $v\to\infty$. Recall, however, that in the \WDW
theory, $b$-representation allowed us to find a more useful
interpretation of the choice of extension in terms of gluing
conditions at $b=\pm b_o$. The similarity of the \WDW and LQC
constraint operators and form of the eigenfunctions when expressed
in terms of $x$ variables (introduced respectively via
\eqref{eq:wdw-x-def} and \eqref{eq:lqc-x-def}) suggests that a
similar interpretation should exist also in LQC. We will now show
that this expectation is correct.

For simplicity we again consider the operator $\Theta'_{\Lambda}$.
Its deficiency functions $\psi^{\pm}\in\defi^{\pm}$ are the weak
solutions to the equation \eqref{eq:lqc-eigenf} with the
eigenfunctions ${\zeta}=\pm i$. Because of symmetry and the global
continuity are of the form
\begin{equation}\label{eq:lqc-defi}
\psi^{\pm}(y) = C \begin{cases}
(1/c_+)(e^{(1\mp i)y/\sqrt{2}} + e^{-(1\mp i)y/\sqrt{2}}) , & |y|<y_o , \\
(1/c_-)(e^{(1\pm i)(\pi-y)/\sqrt{2}} + e^{-(1\pm i)(\pi-y)/\sqrt{2}})
& |y|>y_o , \\
\end{cases}
\end{equation}
where
\begin{subequations}\begin{align}
  c_+ &= e^{(1\mp i)y_o/\sqrt{2}} + e^{-(1\mp i)y_o/\sqrt{2}} , \\
  c_- &= e^{(1\pm i)(\pi-y_o)/\sqrt{2}} + e^{-(1\pm i)(\pi-y_o)/\sqrt{2}} .
\end{align}\end{subequations}
Since all the eigenspaces are non-degenerate, there exists a 1-1
correspondence between these solutions and the appropriate
eigenfunctions of $\Theta_{\Lambda}$ discussed in \cite{kp-posL}.
Since it was shown in \cite{kp-posL} that all eigenfunctions of
$\Theta_{\Lambda}$ are normalizable, it follows that eigenfunctions
\eqref{eq:lqc-defi} are also normalizable. The unitary
transformations between $\defi^+$ and $\defi^-$ and the extended
domains are again given by the exact analogs of
\eqref{eq:wdw-ext-trans} and \eqref{eq:dom-ext} respectively. The
extension-characteristic terms $\psi_{\alpha}$ --analogs of
\eqref{eq:wdw-ext-char}-- are now of the form
\begin{equation}\label{eq:lqc-ext-char}
  \psi_{\alpha}(x) = C' \begin{cases}
                          f(y,y_o,\alpha) , & |y|<y_o , \\
                          f(\pi-y,\pi-y_o,-\alpha) , & |y|>y_o ,
                        \end{cases}
\end{equation}
where $C'\in\compl$ and
\begin{equation}\label{eq:lqc-ext-f}
  f(y,y_o,\alpha) := \f{1}{[\cosh(\sqrt{2}y_o) + \cos(\sqrt{2}y_o)]}
    \,\, \sum_{\rho,\sigma=\pm 1} \hspace{-0.2cm}
    e^{(y+\sigma y_o)\rho/\sqrt{2}}
    \cos\left[\frac{y-\sigma y_o}{\sqrt{2}} + \frac{\rho\alpha}{2}\right] .
\end{equation}

Since $[\partial_x\psi](x=\pm x_o)=0\,\, \forall \psi\in\Dom$,
\,  the extensions are uniquely determined by the parameter
$\beta$,
\begin{equation}\label{eq:lqc-ext-glu}
  \beta(\alpha):=\arctan\left(
  \frac{[\partial^-_y\psi_a](y=y_o)}{[\partial^+_y\psi_a]
  (y=y_o)}\right)\,\,\, \in [0,\pi)\, ,
\end{equation}
that encode the ratios between the left and right derivative of
$\psi_a$ at the gluing point $y=y_o$. With this parametrization, the
choice of self-adjoint extensions can be directly interpreted in
terms of gluing condition at $b=\pm b_o$ also in LQC. (Recall that
in the classical theory, this is where the universe reaches the
infinite volume.) Furthermore for any fixed value of $y_o\in
]0,\pi[$ the function $\beta(\alpha)$ is a bijection of the circle
of radius $1$ onto the circle of radius $1/2$.\, (See
Appendix~\ref{app:bij} for a proof of this assertion). Therefore, as
in the WDW case, $\beta$ is a convenient label for the extensions,
equivalent to $\alpha$, and Eq. \eqref{eq:lqc-ext-glu} succinctly
captures the domain $\Dom_\beta$ of $\Theta_{\Lambda,\beta}$. These
properties will be used in the next subsection to identify the
spectra of particular extensions $\Theta_{\Lambda,\beta}$ and the
corresponding eigenbases in the physical Hilbert spaces.

\subsection{The LQC physical sectors.}

\begin{figure*}[tbh!]
  \begin{center}
    $(a)$\hspace{3.2in}$(b)$
  \end{center}\vspace{-0.2cm}
  \includegraphics[width=3.2in,height=2.4in]{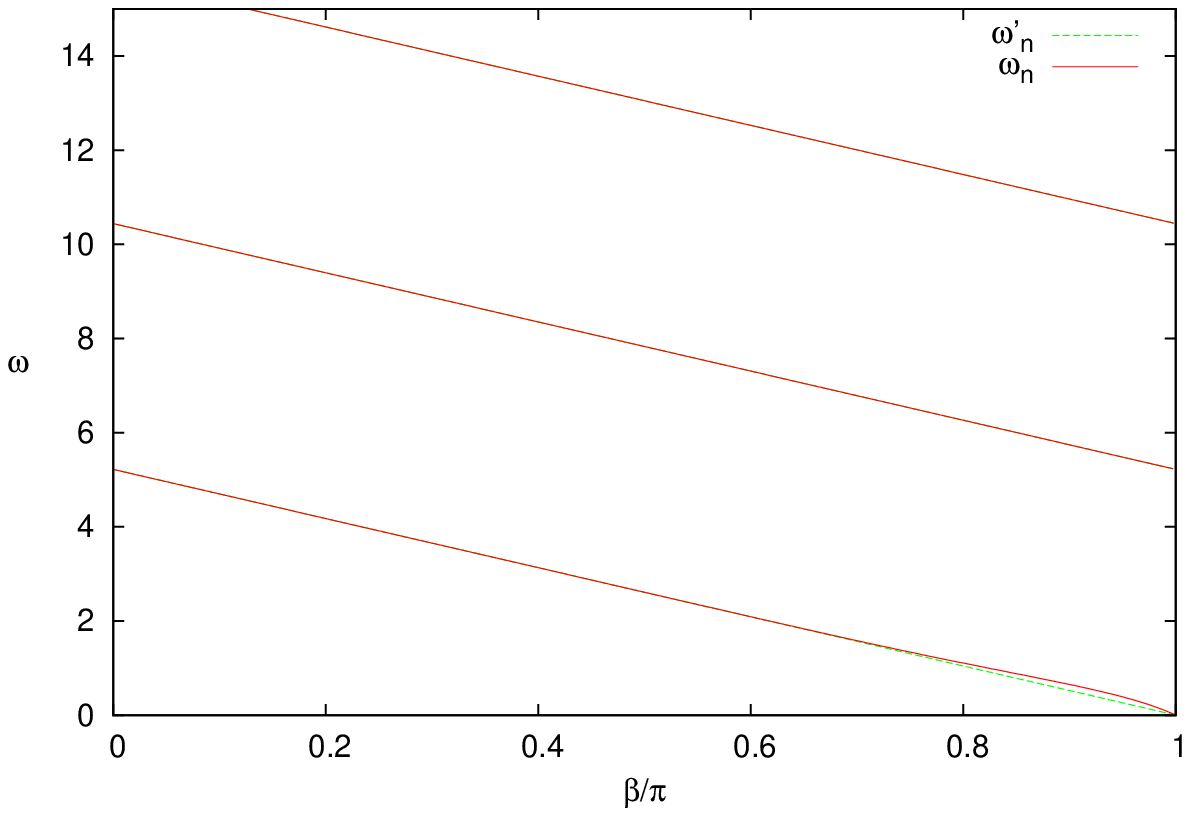}
  \includegraphics[width=3.2in,height=2.4in]{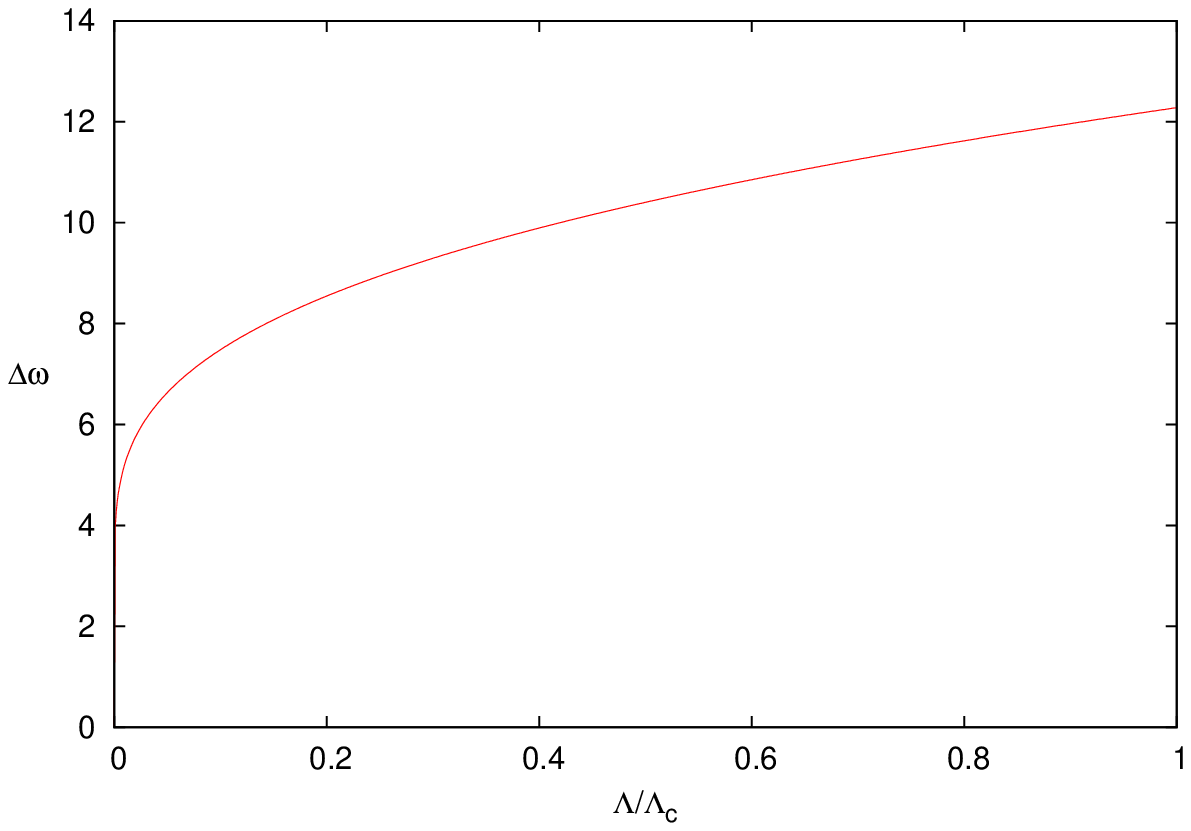}
  \caption{$(a)$ The lowest three eigenvalue $\omega_n$ are plotted
    as functions of the self-adjoint extension parameter $\beta$
    and compared with their approximations $\omega'_n$ obtained by
    neglecting the remnant $O(e^{-2\pi n (\pi-y_o)/y_o})$ in
    \eqref{eq:sp-approx}, for $\Lambda= 0.01 \Lambda_c$.
    $(b)$ the frequency gap $\Delta\omega$ is plotted as a function of
    cosmological constant.
  }
  \label{fig:omegas}
\end{figure*}

We can now fix any one self-adjoint extension
$\Theta_{\Lambda,\beta}$ of $\Theta_{\Lambda}$ in the quantum
Hamiltonian constraint \eqref{eq:constr-quant} and perform group
averaging to obtain the physical Hilbert space
$\Hil^{\phy}_{\beta}$. Because the spectra of
$\Theta_{\Lambda,\beta}$ are discrete \cite{kp-posL}, the physical
states are of the form
\begin{equation}\label{eq:lqc-phys-state}
  \Psi(x,\phi) = \sum_{n=0}^{\infty} \tilde{\Psi}_n \, e_{\beta, n}(x) \,
  e^{i\omega_{\beta, n}(\phi-\phi_o)} ,
\end{equation}
where the spectral profiles are square-summable sequences
$\tilde{\Psi}_n$, and $e_{\beta, n}$ are the normalized
eigenfunctions of $\Theta_{\Lambda,\beta}$ (belonging to
$\Dom_{\beta}$), with eigenvalues $\omega^2_{\beta, n}$.
Alternately, $e_{\beta, n}$ are eigenfunctions of the operator
$\Theta'_{\Lambda,\beta}$ with eigenvalues $k_n^2$, related to
$\omega_n$ via $\omega_n=(\sqrt{12\pi G}\,\pi/x_M)\, k_n$. They
satisfy Eq. \eqref{eq:lqc-ext-glu} and their functional form is
given by
\begin{equation} \label{eigenfunctions}
  e_{\beta, n}(x) = N_{\beta, n}
  \begin{cases}
    \cosh[k_n(\pi-y_o)]\,\cos(k_n y) , & |y|<y_o , \\
    \cos[k_n y_o]\,\cosh[k_n(\pi-y)] , & |y|>y_o ,
  \end{cases}
\end{equation}
where $N_{\beta, n}$ is a normalization factor and the eigenvalues
$k_{\beta, n}$ are determined by the condition
\begin{equation}\label{eq:lqc-spectrum}
  \tan(k_n y_o) + \tanh[k_n(\pi-y_o)]\, \tan(\beta) = 0 .
\end{equation}
The form of \eqref{eq:lqc-spectrum} implies, in particular,
that the spectra of $\Theta_{\Lambda,\beta}$ for different
values of $\beta$ are \emph{disjoint}. Furthermore the lowest
non-negative eigenvalue $\omega^2_{\beta, 0}$ is an invertible
function of $\beta$, whence the physical Hilbert spaces
$\Hil^{\phy}_{\beta}$ corresponding to each extension are in
fact \emph{different} subspaces of $\Hil_{\gr}$.

Although the eigenvalues $k_n^2$ are provided only implicitly,
the relation \eqref{eq:lqc-spectrum} allows us to determine
their asymptotic behavior for large $n$ explicitly:
\begin{equation}\label{eq:sp-approx}
  k_n = (n\pi-\beta)/y_o + O(e^{-2\pi n (\pi-y_o)/y_o}) .
\end{equation}
Consequently, as $n$ grows, the distribution of $k_n$ quickly
approaches the uniform one with the separation $\Delta k =
\pi/y_o$. This property transfers directly to
$\Theta_{\Lambda}$, where
\begin{equation}
  \lim_{n\to\infty} [\omega_{n+1}-\omega_n] = \Delta\omega
  := \sqrt{12\pi G}\,\pi^2/x_o ,
\end{equation}
Although the distribution of lower frequencies is not quite
uniform, it approaches uniformity extremely fast (see
Fig.~\ref{fig:omegas}a), and the asymptotic separation
$\Delta\omega$ depends only on the value of the cosmological
constant (see Fig.~\ref{fig:omegas}b). In particular
\begin{equation}
  \lim_{\Lambda\to 0} \Delta\omega = 0  \quad {\rm and} \quad
  \lim_{\Lambda\to \Lambda_c} \Delta\omega = 2\pi\sqrt{12\pi G} .
\end{equation}

The form of the quantum Hamiltonian constraint
\eqref{eq:constr-quant} and of the physical states
\eqref{eq:lqc-phys-state} allows us again to use the scalar field as
an internal time and regard the square root of the positive part of
$\Theta_{\Lambda,\beta}$ as the generator of time evolution on the
subspace $\Hil^{\phy}_{\beta} = P^+_{\beta} \Hil_{\gr}$ of
$\Hil_{\gr}$:
\begin{equation}\label{eq:lqc-evo}
  \Psi(x,\phi_o)\,\, \mapsto\,\, \Psi(x,\phi)
  = e^{i(\phi-\phi_o)\sqrt{|\Theta_{\Lambda,\beta}|}}\,\,\Psi(x,\phi_o)
  \qquad \forall\,\, \Psi(x,\phi) \in P^+_{\beta}\Hil_{\gr} ,
\end{equation}
where, as before, $P^+_{\beta}$ is a projection onto the positive
part of the spectrum of $\Theta_{\Lambda,\beta}$.

To obtain the physical consequences of this evolution, we will use
observables $|\hat{\theta}|_{\phi}$ and $\hat{\rho}_{\rm
tot}|_{\phi}$ in exact analogy with Eqs. \eqref{eq:wdw-theta-def}
and \eqref{eq:wdw-rho-def} of the \WDW theory. However, in contrast
to the \WDW theory, we cannot define $|\hat{b}|_{\phi}$ since the
operator $\partial_v$ fails to exist in LQC;  we can only define its
bounded, periodic functions such as $\sin(\hat{b})|_{\phi}$.

\subsection{LQC dynamics}\label{sec:lqc-dyn}

To facilitate comparison with the \WDW theory, we will use states
closely resembling the Gaussians used in Sec.~\ref{sec:wdw-dynamics}
peaked about large $\omega^\star$:
\begin{equation} \label{profile}
  \tilde{\Psi}_n = e^{-\f{(\omega_n-\omega^\star)^2}{2\sigma^2}}\,\,
   e^{-i\omega_n\phi^\star}
\end{equation}
where $\phi^\star$ is given by \eqref{eq:phistar}. we can then
repeat the procedure used in the WDW theory to carry out numerical
simulations. Specifically, these computations were performed in the
following steps:

\begin{enumerate}[(i)]
\item First, the spectrum $\Sp(\Theta_{\Lambda,\beta})$ was
    found from \eqref{eq:lqc-spectrum}. In all cases considered,
    the approximation provided by the analytical expression
    \eqref{eq:sp-approx} turned out to be excellent.
\item Given any eigenvalue in $\Sp(\Theta_{\Lambda,\beta})$ the
    corresponding eigenfunction $\psi_{\beta,n}(v)$ was found in
    the domain $v\in\{4n:\ n=1 \dots N_1\}$ by solving the
    iterative difference equation
    \begin{equation}
      \omega^2_n\,\psi_{\beta,n}(v) = [\Theta_{\Lambda,\beta}\psi_{\beta,n}](v)
    \end{equation}
    with initial data $\psi_{\beta,n}(v=4)=(-1)^n$. In various
    simulations, $N_1$ ranged from $10^5$ to $2.5\times 10^5$.
    Using the expression \eqref{eq:ip-gr} of the scalar product
    in the $v$-representation, the norm of $\psi_{\beta,n}(v)$
    was calculated using a polynomial extrapolation. The
    normalized eigenfunctions $\bar{e}_{\beta,n}(v)$ were then
    used together with the profile coefficients $\tilde\Psi_n$
    of Eq. \eqref{profile} to obtain the physical state
    $\Psi(v,\phi)$ in the $v$-representation:
    \be \Psi(v,\phi)= \sum_{n=0}^{\infty} \tilde{\Psi}_n \,
    \bar{e}_{\beta, n}(v) \, e^{i\omega_{\beta,
    n}(\phi-\phi_o)}\ee
    where the sum extends over those $n$ for which $\omega_n \in
    [\omega^\star-7\sigma,\omega^\star+7\sigma]$. This
    state is defined on points $v\in\{4n:\ n=1 \dots N_1\}$ and
    is simply the transform of the state in Eq.
    \eqref{eq:lqc-phys-state} to the $v$-representation. A key
    difference between the \WDW theory is that, once the
    eigenvalues are obtained using \eqref{eq:lqc-spectrum}, we
    can work directly in the $v$ representation because, unlike
    in the \WDW theory, the LQC eigenfunctions $\bar{e}_{\beta,
    n}(v)$ in the $v$ representation can be easily handled
    numerically. (Recall that in the \WDW theory, the
    corresponding eigenfunctions are Bessel functions of
    imaginary order which are difficult to deal with
    numerically.)
\item Expectation values and dispersions of the relational Dirac
    observables, $|\hat{\theta}|_{\phi}$, $\hat{\rho}_{\rm
    tot}|_{\phi}$ were computed directly in the $v$
    representation using the expressions \eqref{ops} of
    $\hat{v}$ and the shift operator $\hat{\cal N}_\mu$. In
    particular, when the operator $\hat{\rho}_{\rm tot}$ is
    factor ordered as in the \WDW theory, it becomes a simple
    combination of the shift operators:
\begin{equation}
  \hat{\rho}_{\rm tot} = \frac{3}{8\pi G \lambda^2 \gamma^2} [\widehat{\sin(b)}]^2\,.
\end{equation}
Note that the expression has the same form as
\ref{eq:wdw-rho-def}) in the \WDW theory, but the \WDW operator
$\hat{b}^2$ is now replaced by the bounded operator
$\widehat{\sin(b)}^2$.
\end{enumerate}

\begin{figure*}[tbh!]
  \begin{center}
    $(a)$\hspace{3.2in}$(b)$
  \end{center}\vspace{-0.2cm}
  \includegraphics[width=3.2in,height=2.4in]{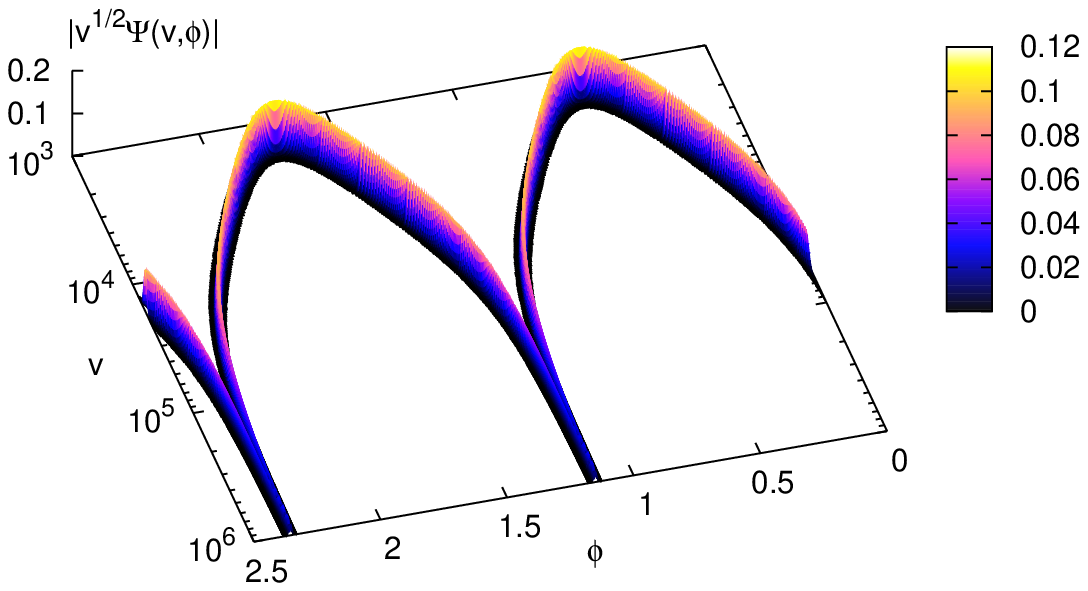}
  \includegraphics[width=3.2in,height=2.4in]{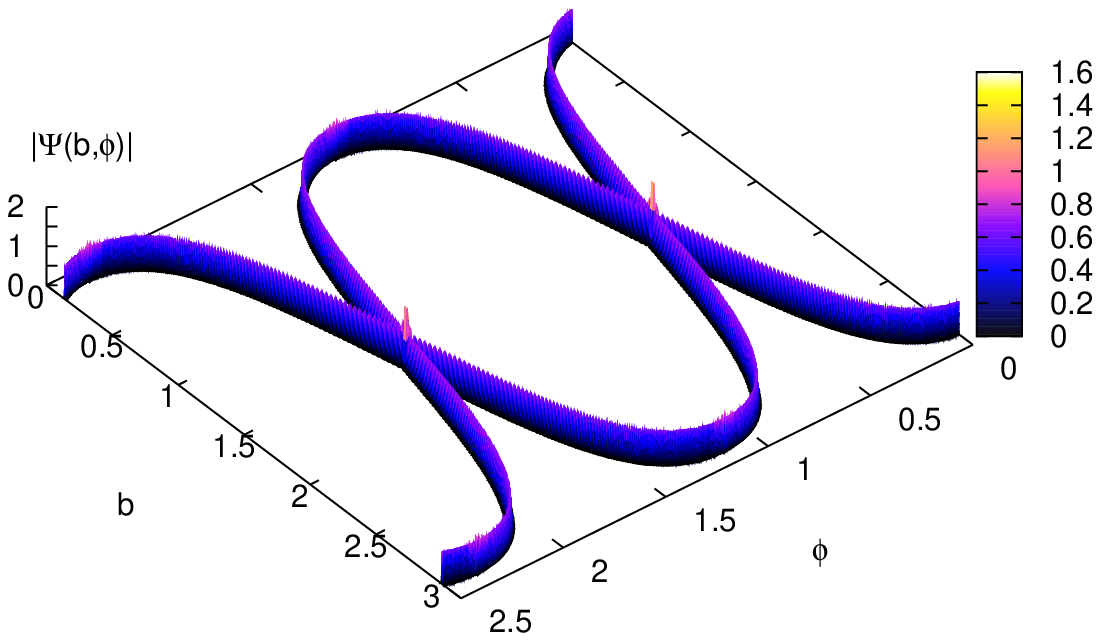}
  \begin{center}
    $(c)$\hspace{3.2in}$(d)$
  \end{center}\vspace{-0.2cm}
  \includegraphics[width=3.2in]{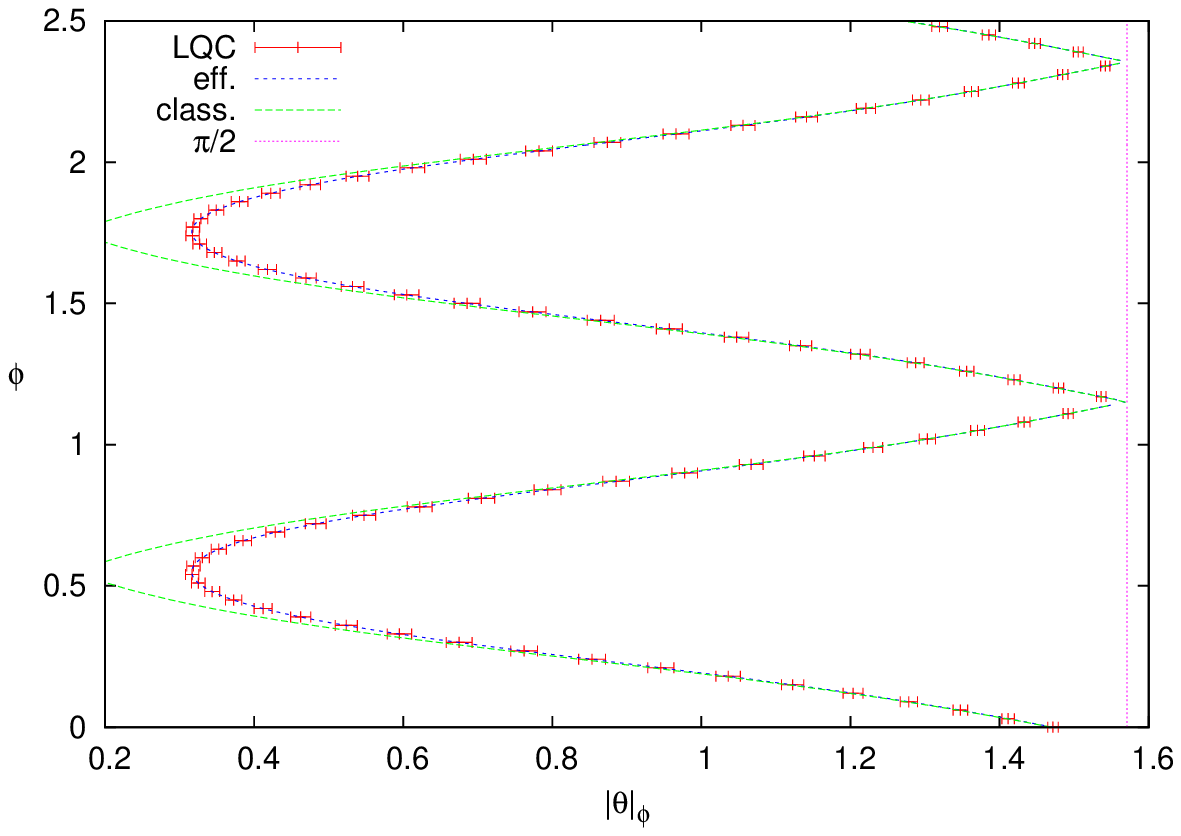}
  \includegraphics[width=3.2in]{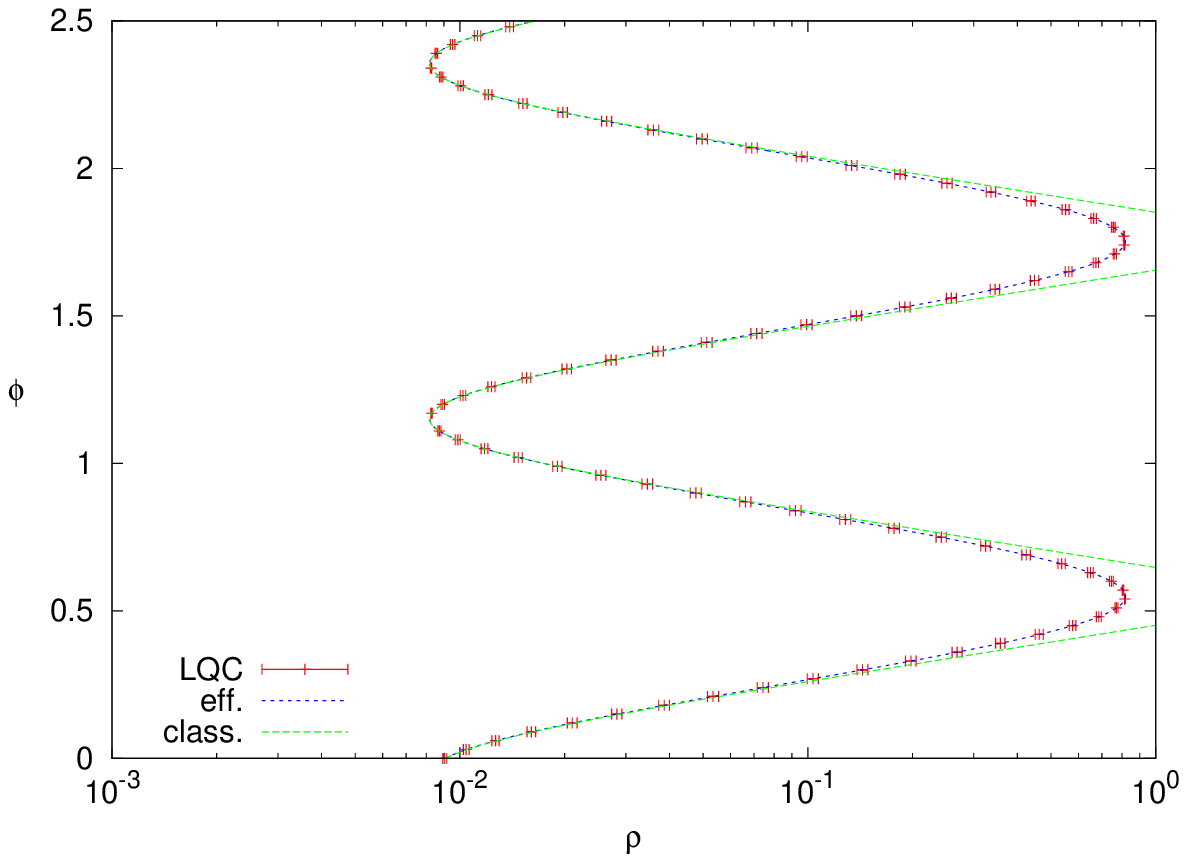}
  \caption{LQC dynamics. Figures (a) and (b) show the
    evolution of the (absolute value of the) physical wave function
    in the $v$-$\phi$ and
    $b$-$\phi$ planes respectively. Figures (c) and (d) compare the
    expectation values of the Dirac observables $|\hat{\theta}|_\phi$
    and $\hat{\rho}_{\rm tot}|_\phi$ with those in the effective and
    classical theories. Away from the Planck regime there is excellent
    agreement with (the analytical extensions of) solutions in general
    relativity. But in the Planck regime (left ends of figures (c) and
    (d)) there is a very large departure from the classical behavior
    because of quantum geometry effects. However, the effective
    trajectories capture the quantum evolution very well even in the
    Planck regime. In these simulations, the initial state was a
    gaussian  peaked at $p_{\phi}^\star=5\times 10^3$ and
    $|\hat{\theta}|^\star_{\phi=0} = \arctan(10)$ with the relative
    spread $\Delta p_{\phi}/p_{\phi}=0.03$. The self-adjoint extension
    corresponded to $\beta=0$, the constant $K$ in the definition of
    $\theta$ was set to $K= 5\times 10^3$ and the cosmological constant
    was $\Lambda = 0.01 \Lambda_c \approx 0.2\lPl^{-2}$.}
  \label{fig:lqc-dynamics}
\end{figure*}

As in the \WDW theory, a large number of simulations were performed
by varying parameters of the initial state and the self-adjoint
extension used in quantum dynamics. Fig.~\ref{fig:lqc-dynamics}
illustrates the results of a typical simulation. The qualitative
behavior is the same as that in the $\Lambda < 0$ case \cite{bp}.

\begin{itemize}
\item States under consideration remain sharply peaked over a
    large number of `epochs', where each `epoch' is
    characterized as the evolution between consecutive quantum
    bounces. Thus, in each epoch the universe starts out with
    the maximum but finite total density $\rcr$, expands out till
    the matter density  $\rho|_{\phi}$ vanishes and then recollapses,
    the density again reaching $\rcr$ at the end of the epoch.
\item Expectation values of the Dirac observables
    $|\hat{\theta}|_\phi$ and $\hat{\rho}_{\rm tot}|_\phi$ are
    well approximated by the classical effective dynamics
    discussed in Appendix~\ref{app:eff} \emph{throughout} the
    evolution, including the quantum bounces. The difference
    between the two is much smaller than the dispersions of the
    corresponding Dirac observables.
\item For $\Lambda\ll \Lambda_c := 3/\gamma^2\lambda^2$, quantum
    dynamics is well-approximated by the (analytically extended)
    classical general relativity trajectory so long as the
    scalar field energy density is small compared to the Planck
    scale, $\rho|_\phi\ll\rho_{\Pl}$.
\item However, as $\rho|_{\rm tot} = \rho|_\phi+\Lambda/(8\pi
    G)$ approaches the Planck scale, an effective repulsive
    force due to quantum geometry starts to dominate. It soon
    overwhelms the classical gravitational attraction and forces
    a quantum  bounce at $\rho_{\rm tot}=\rcr$. As a result, the
    contracting quantum universe bounces back into an expanding
    quantum universe. As noted above, the wave functions remain
    sharply peaked even during the bounce and expectation values
    of Dirac observables follow those given by the effective
    equations even in the Planck regime.
\item As in the \WDW theory, quantum evolution is not unique
    because it depends on the choice of the self-adjoint
    extension $\Theta_{\Lambda,\beta}$ of $\Theta_{\Lambda}$.
    Each extensions corresponding to the choice of a specific
    reflective boundary condition, but at $\rho_{\rm tot} =
    \Lambda/(8\pi G)$, and \emph{not in the Planck regime near}
    $\rho_{\rm tot} \approx \rcr$.
\item For any chosen self-adjoint extension, as in the \WDW
    theory, the expanding universe recollapses after reaching
    zero matter density. However, unlike in the \WDW theory,
    the big bang and the big crunch is replaced by quantum
    bounces. Therefore, we now have a nearly periodic evolution
    in LQC admitting an infinite chain of the bounces at the
    Planck energy density and recollapses at zero matter density
    The duration of each `epoch' is given by
    \begin{equation}\label{eq:lqc-period}
      \Delta\phi \,\approx \,\f{4y_o\Delta k}{\Delta\omega}\, \approx\,
      \f{2x_o}{\sqrt{3\pi G}} .
    \end{equation}
    to an excellent approximation.
\item However, the distribution of $\omega_n$ is not exactly
    uniform. Consequently, quantum states slowly disperse from
    one cycle to the next. But for large $\omega_n$ deviations
    from uniformity decay  much faster those in the case of
    $\Lambda < 0$ considered in \cite{bp}. Therefore the rate of
    dispersing is much slower than the already low rate found in
    \cite{bp}.
\item As in the \WDW theory, the dependence on the choice of the
    self-adjoint extension is surprisingly weak for the states
    considered here. For generic states, results of the S-matrix
    discussion of \cite{kp-posL} show that the dependence on
    $\beta$ of expectation values of Dirac observables is
    negligible compared to the dispersion of the corresponding
    quantities. The dependence of dispersions on $\beta$ is also
    very small compared to the dispersion $\delta_{\ln |k|}$ in
    the scalar field momentum (which is a constant of motion).
\end{itemize}

\section{Discussion}
\label{s5}

In this paper, we analyzed in detail the \WDW theory and LQC of the
k=0, $\Lambda>0$ FLRW model along the lines of the treatment of the
$\Lambda <0$ case of \cite{bp}, thereby completing the program
outlined in an Appendix of \cite{aps3}. As in the $\Lambda \le 0$
cases, the scalar field can be used as a global clock, providing us
with a natural notion of relational time both in the classical and
quantum theories. However, interestingly, there is a key difference
in the physically most interesting case, that of $\Lambda >0$: In
classical general relativity, solutions that start with infinite
matter density $\rho$ at the big bang at time $\phi = -\infty$
expand out and now achieve $\rho=0$ at some \emph{finite} value
$\phi_o$ of internal time (when the volume $v$ of the fiducial cell
$\mathcal{C}$ becomes infinite). Thus, in the $\rho$-$\phi$ plane
each of these dynamical trajectories starts out at $\phi = -\infty$
but ends at $\phi=\phi_o$. But it can be \emph{analytically}
extended beyond $\phi=\phi_o$ and the extension represents a
universe which starts out with zero matter density at $\phi=\phi_o$
but contracts, ending in a big crunch singularity at $\phi=\infty$.
>From the relational time perspective, then, one is led to regard the
two branches as providing a single dynamical trajectory because it
is artificial to simply end dynamics at a finite value of time. Now,
in non-relativistic mechanics if the potential is negative and steep
the particle may roll off to infinity in a finite amount of time. In
that case, one has to choose from a one (or more) parameter family
of boundary conditions at infinity to continue dynamics beyond that
time. In the present case, by contrast, we did not have to resort to
making a choice because the most interesting Dirac observable,
$\rho|_\phi$, is analytic in $\phi$.

Nonetheless, the fact that the universe expands out to $\rho=0$ at a
finite value of the relational time introduces ambiguities in the
\emph{quantum} evolution: The operator $\Theta_\Lambda$ which
generates dynamics with respect to $\phi$ now fails to be
essentially self-adjoint. Before discussing this point in detail,
let us first note two aspects of this phenomenon. First, it is not a
peculiarity of LQC; it occurs also in the \WDW theory. Second, in
both cases, the lack of essential self-adjointness is related
directly with the behavior of the system at large $v$ and low matter
density $\rho$; its origin does \emph{not} lie in the Planck scale
physics. In both quantum theories, the dynamical operator admits a
one parameter family of self-adjoint extensions. For a general
system, different choices of extensions can give rise to very
different dynamics. However, in this model the results are
surprisingly robust with respect to this choice. Not only is the
qualitative behavior of dynamics the same, but the differences in
the dynamics of the expectation values of the most interesting Dirac
observables in theories resulting from two different extensions are
smaller than their dispersions in any one theory, even for general
states. Furthermore, numerical simulations show that, irrespective
of the choice of extension, quantum states which are semi-classical
in the low curvature (or low total density $\rho_{\rm tot}$) regime
remain sharply peaked at the \emph{extended} classical trajectory in
the low curvature regime both in the \WDW theory and LQC. It is
tempting to conjecture that this robustness of quantum dynamics is
related to the fact that we did not have to choose a boundary
condition at $\phi=\phi_o$ to extend the classical $\rho-\phi$
trajectory. In the remainder of this section, most of our discussion
on the behavior of wave functions will refer to these states.

As in the $\Lambda \le 0$ cases, there is a \emph{pronounced}
difference between the quantum dynamics of the two theories in
the Planck regime. In the \WDW theory, the wave function simply
follows the extended classical trajectory into the big-bang and
the big-crunch singularities. In LQC, by contrast, while these
states remain peaked at the classical trajectory so long as the
curvature (or $\rho_{\rm tot}$) is low compared to the Planck
scale, there is a dramatic departure in the Planck regime.
There is again a new repulsive force with origin in the quantum
geometry that overwhelms classical gravity and cases a quantum
bounce. Again, the numerical simulations show that, although
the force is so strong in the Planck regime, it dies very
quickly and becomes negligible once $\rho_{\rm tot}$ falls
below  $10^{-2}- 10^{-3}$ Planck density. In LQC then, even
though we are in the k=0 case, we are led to a scenario that is
approximately cyclic. As in the k=1 LQC models, the quantum
evolution spans an infinite number of epochs. In each epoch the
universe begins with a quantum bounce where $\rho_{\rm tot}
\approx 0.41 \rho_{\rm Pl}$, expands out till $\rho_{\rm tot} =
\Lambda/8\pi G$ and then undergoes a collapse till it reaches
another quantum bounce. For states under consideration,
dynamics is nearly periodic.

How does this dynamics appear in the space-time picture? Let us
begin with the classical theory and consider a solution in which the
universe starts out with a big-bang at $\phi=-\infty$. It expands
out to $\scri^+$ ---which is space-like for $\Lambda>0$--- where the
matter density vanishes and $\rho_{\rm tot} = \Lambda/8\pi G$. In
terms of the physical metric, this space-time is future complete.
However, the extended phase space trajectory analytically continues
the space-time geometry across $\scri^+$, effectively gluing it with
$\scri^-$ of a contracting solution.%
\footnote{The detailed gluing procedure will involve a
  conformal completion along the lines of \cite{ar} where the
  normal component to $\scri$ of the metric is rescaled by a
  different power of the conformal factor than the tangential
  one.}
Quantum states under consideration remain peaked at these extended
space-time geometries across $\scri$. An extension is but to be
expected both in the \WDW theory and in LQC: quantum evolution in
the internal time $\phi$ is \emph{unitary} and $\phi$ achieves a
finite value $\phi_o$ at $\scri$ of the given classical solution,
unitary evolution could not just stop there. What is interesting is
that, \emph{irrespective of the choice of the self-adjoint
extension}, the state remains sharply peaked on the analytically
extended geometry.

This extension, and the ensuing nearly cyclic scenario has some
similarities with Penrose's recent proposal of a cyclic
conformal cosmology \cite{rp}. However, there are also key
differences. In our case, $\scri^+$ of the expanding branch is
glued to the $\scri^-$ of the contracting branch; not to the
big-bang singularity of the next `aeon'. More importantly,
quantum geometry effects are crucial in LQC. In particular
$\hbar$ appears in the denominator of the expression of the
maximum density $\rcr$ whence, as one would expect, $\rcr$
would diverge in the classical limit $\hbar \to 0$. Therefore,
quantum effects and a non-zero $\hbar$ play an essential role
in the approximately cyclic scenario of LQC. By contrast, a
central feature of the cyclic conformal cosmology paradigm is
that, although one does have unboundedly large curvatures,
$\hbar$ plays no role at all in this regime.

Use of the scalar field as a relational time variable played a
key role throughout our analysis, both in the classical and
quantum theory. What would have happened if we had made some
other choice? In a recent analysis \cite{hp-qg,hp-qc}
non-rotating dust has been used in place of the scalar field.
In this case, the expression (\ref{eq:constr-quant}) of the
gravitational part $\Theta_\Lambda$ of the Hamiltonian
constraint is modified because the lapse is now tailored to
proper time. In particular, the coefficient of $\Lambda$ is now
\emph{linear} rather than \emph{quadratic} in $v$.
Consequently, the analog of $\Theta_\Lambda$ is now essentially
self-adjoint and the LQC evolution resembles that in the
$\Lambda=0$ case \cite{aps3}: The universe starts out with
infinite volume in the distant past, collapses, undergoes a
quantum bounce and then expands out to infinite volume.
However, in the Planck regime, quantum matter should be
described using quantum field theory and for all standard
quantum fields the kinetic term in the Hamiltonian is quadratic
in momenta. Therefore the specific feature that simplifies the
mathematics in the case of dust is no longer available and the
overall situation is then the same as that in the case of the
scalar field.

But what if we return to using scalar field as matter source
but let a geometric variable be the relational time? An obvious
choice is volume. But in LQC, volume fails to be single valued
making it difficult to introduce the `time-dependent'
relational Dirac observables ---such as the matter density
operator $\hat\rho|_v$--- especially near the bounce. On the
other hand, the conjugate variable $b$ is single valued both on
classical and effective trajectories and, as the form of the
Hamiltonian constraint suggests, the function $y$ it determines
is a possible candidate for relational time. However, it
appears that the evolution would then be unambiguously unitary
only for $|y| >y_o$. Furthermore, defining the physical state
would require specification of the initial data at $y=y_o$ but
the theory does not provide any selection principle for this
task. These difficulties with $v$ and $b$ could well be
surmountable with new ideas and more careful analysis. But as
of now they seem to be more serious handicaps than the
complications associated with the use of the scalar field as
relational time we encountered in this paper.

Could one perhaps retain scalar field as the matter source but
set lapse $N=1$ in the quantum constraint and use a `timeless
framework' in LQC? Results of \cite{klp1,klp2} imply that we
would have been led to a theory that is mathematically free of
ambiguities associated with self-adjoint extensions. However,
as we now explain, this theory is difficult to interpret and it
is unclear whether it is physically viable. We will conclude
our discussion with a detailed elaboration of this point.

With lapse $N=1$, the quantum Hamiltonian constraint has the
form \, $\hat{C}= B(v) \otimes \partial_\phi^2 +
\hat{C}_{\gr}\otimes \id$ where $C_{\gr}$ is the gravitational
part of the Hamiltonian constraint and $B(v) \sim 1/v$ for
large values of $v$ (see, e.g., \cite{aps3}). This total
constraint operator $\hat{C}$ has been shown to be essentially
self-adjoint \cite{kp-posL} and for simplicity we will denote
its self-adjoint extension also by $\hat{C}$. One can therefore
use it directly for group averaging and construct the physical
Hilbert space $\Hil^{\phy}$ in the `timeless framework' without
recourse to deparametrization. Then, although states can be
represented as wave functions $\Psi(v,\phi)$, the physical
scalar product is no longer given by an integral over $v$ at a
fixed value of $\phi$. What is the relation between this
$\Hil^{\phy}$ and the Hilbert spaces $\Hil^{\phy}_{\beta}$
associated with self-adjoint extensions
$\Theta_{\Lambda,\beta}$ we constructed in this paper? It turns
out \cite{klp2} that $\Hil^{\phy}$ is huge; it is given by the
\emph{direct integral} of all $\Hil^{\phy}_{\beta}$;\,\,
$\Hil^{\phy}= \int_{\oplus\, I} \Hil^{\phy}_{\beta} \,
\rd\beta$, where $I$ is the interval $(0, \pi)$.

We will briefly discuss a simple example ---due to Wojciech
Kami\'nski in \cite{k-pres}--- to illustrate the relation
between these Hilbert spaces. Consider a 2-dimensional strip,
$M= \mathbb{R} \times [0,1]$, with coordinates $\phi \in
\mathbb{R}$ and $x\in [0,1]$ and a constraint thereon in the
form of the Schr\"odinger equation $(-i\partial_\phi +
i\partial_x)\Psi(x,\phi) =0$ (with a first-order Hamiltonian).
As in LQC, $\phi$ plays the role of time while $x$ is to be
thought of as the analog of the compactified volume coordinate
$\theta$. The operator $\Theta_x := -i\partial_x$ fails to be
essentially self-adjoint on the closed interval $[0,1]$; it
admits a 1-parameter family of self-adjoint extensions
$\Theta_{x,\beta}$, labeled by $\beta \in [0, 2\pi)$ (with
domain $\Dom_\beta$ given by wave functions $\psi(x)$
satisfying $\psi(1) = e^{i\beta}\psi(0)$). For each extension,
we can construct a physical Hilbert space $\Hil^{\phy}_\beta$
in the standard manner from solutions to the quantum
constraint: $\Psi(x,\phi) \in \Hil^{\phy}_\beta$ if and only if
$-i\partial_\phi \Psi(x,\phi) = \Theta_{x,\beta}\,
\Psi(x,\phi)$. Next, let us define $\Hil := \int_{\oplus\, I}
\Hil^{\phy}_{\beta} \, \rd\beta$, with $I= (0,2\pi)$. ($\Hil$
is analogous to $\Hil^{\phy}$ obtained by group averaging in
\cite{klp2}). Every $\Psi_\beta(x,\phi) \in \Hil^{\phy}_\beta$
can be expanded as
\be \Psi_\beta(x,\phi) = \sum_{n=-\infty}^{\infty}
\tilde{\psi}_{\beta, n}\,\, e^{i(2\pi n+\beta) x}\,\,
e^{ik\phi} \ee
where $k:=2\pi n+\beta$. On the other hand $\Psi(x,\phi) \in \Hil$
has the form
\begin{equation}\begin{split}
  \Psi(x,\phi) &= \int_0^{2\pi}\!\! \rd\beta\, \Psi_\beta(x,\phi)
  = \sum_{n=-\infty}^{\infty} \, \int_0^{2\pi}\!\rd \beta\,
  \tilde{\psi}_{\beta,n}\,
  e^{i(2\pi n +\beta) x}\, e^{ik\phi}  \\
  &= \int_{-\infty}^{\infty}\!\! \rd k\, \tilde\Psi(k)\,
  e^{ikx + ik\phi}\, ,
\end{split}\end{equation}
where we have set $k = 2\pi n +\beta$ as above. The norms in
$\Hil^{\phy}$ are given by:
\be \|\Psi(x,\phi)\|^2 = \int_{-\infty}^{\infty}\!\! \rd k\,
|\tilde{\Psi}(k)|^2\, . \ee
Note that $\tilde{\Psi}(k)$ has support on the entire $k$-axis and
furthermore, the norm of $\Psi(x,\phi)$ also involves the integral
of $|\tilde{\Psi}(k)|^2$ over the entire $k$ axis. Therefore, while
elements $\Psi_\beta(x,\phi)$ of any one $\Hil^{\phy}_\beta$ are
restricted to have support only on the physical configuration space
$M = \mathbb{R}\times [0,1]$, $\Psi(x,\phi)$ in $\Hil^{\phy}$ are
allowed to be non-zero all along the $x$-axis, even though points
outside the $x$-interval $[0,1]$ have no physical interpretation in
the model under consideration. In other words, although we
restricted ourselves only to a `strip' $M= \mathbb{R}\times [0,1]$
of the Minkowski space in defining the classical system, in effect
$\Hil^{\phy}$ describes a system on the entire 2-dimensional
Minkowski space $\mathbb{R}^2$.
\footnote{Note that, since the basis functions $e^{i(2\pi
n+\beta)x}$ can be extended analytically to the entire $x$-axis, the
quantum constraint can be uniquely extended to the full Minkowski
space $\mathbb{R}^2$.}
To summarize, for the particle on the strip $M$, one would
physically expect that the quantum theory should be formulated
just on $M$ since values of $x$ outside $[0,1]$ have no
physical meaning. This expectation is borne out if one works
with any one self-adjoint extension $\Theta_{x,\beta}$ of
$\Theta_{x}$ but not if one works with the direct integral
$\Hil^{\phy}$ of all the resulting Hilbert spaces.

In LQC, the situation is analogous. Working with a specific
self-adjoint extension allows us to remain in the interval $[0,
\infty]$ of the $|v|$-axis, introduce Dirac observables and
track their evolution in the internal time $\phi$ and compare
it with classical trajectories. On the other hand, working in
the timeless framework in effect requires us to extend the
$v$-axis beyond $v=\infty$ and this extension is difficult to
interpret physically.%
\footnote{Furthermore, unlike in the simple example discussed
above, here one cannot extend the constraint operator uniquely
even if one expresses the basis vectors as functions of the
`compactified volume' coordinate $\theta$ of
\eqref{eq:p-theta}, as these functions are not analytic in
$\theta=\pi/2$.}
For the same reason, while one can introduce Dirac observables
also in the timeless framework, we cannot ask for their
`evolution' and it is difficult to compare predictions of the
quantum theory with those of the classical. In particular,
while the choice $N=1$ of the lapse yields evolution in proper
time in the classical theory, unfortunately this interpretation
does not extend to the quantum theory in a simple way. Thus,
while at first it seems mathematically natural to work with the
lapse $N=1$ because the full constraint is then essentially
self-adjoint, the resulting Hilbert space $\Hil^{\phy}$ appears
to be simply too large to be physically viable in the above
context.

Finally, our analysis brought out an unexpected robustness of the
quantum evolution with respect to the choice of self-adjoint
extensions $\Theta_{\Lambda,\beta}$. As we noted above, this may be
related to the fact that, in the physical sector associated with
\emph{any} self-adjoint extension, under unitary evolution quantum
states of interest follow the natural and unambiguous analytic
extension of classical trajectories. Is this perhaps a special case
of as yet unknown general result? Is the lack of sensitivity of the
quantum evolution on the choice of analytic extensions have its
origin in some special features of the classical evolution?

\section*{Acknowledgments}

We would like to thank Wojciech Kami\'nski and Parampreet Singh
for discussions. This work was supported in part by the NSF
grants PHY0854743 and PHY1068743, and by the Natural Sciences
and Engineering Research Council of Canada. The numerical
analysis has been performed with use of the numerical LQC
library presently developed by T.~Paw{\l}owski and J.~Olmedo.

\appendix

\section{Bijectivity of $\beta(\alpha)$}\label{app:bij}

In section \ref{sec:LQC-Theta-prop} we began by labeling the self-adjoint
extensions of $\Theta_\Lambda$ by a parameter $\alpha$ and then
switched to a more convenient parameter $\beta$. In this appendix,
we will show that the map $\alpha \to \beta$ is bijective, i.e. that
the $\beta$ parametrization used to denote self-adjoint extensions
as $\theta_{\Lambda, \beta}$ is viable.

The definition of $\beta$ and the periodicity
$\beta(\alpha)=\beta(\alpha+\pi)$ ---a direct consequence of its
form given by \eqref{eq:lqc-ext-char}, \eqref{eq:lqc-ext-f} and
\eqref{eq:lqc-ext-glu}--- imply that $\beta(\alpha)$ is a well
defined continuous function mapping from a circle of radius $1$ to a
circle of radius $1/2$. Therefore to establish the desired bijective
property it suffices to prove that, for any given $y_o\in ]0,\pi[$,
the derivative of $\beta$ with respect to $\alpha$ is bounded and
isolated from zero.

Let us consider first the function $X(y_o,\alpha) :=
\tan(\beta)$ which can be decomposed as follows
\begin{equation}
  X(y_o,\alpha) = \frac{f_1(y_o) g_1(y_o,\alpha)}{f_2(y_o) g_2(y_o,\alpha)} ,
\end{equation}
where
\begin{subequations}\begin{align}
  f_1 &= \cosh(\sqrt{2}(\pi-y_o))  + \cos(\sqrt{2}(\pi-y_o)) , \\
  f_2 &= \cosh(\sqrt{2}y_o) + \cos(\sqrt{2}y_o) ,
\end{align}\end{subequations}
and
\begin{subequations}\begin{align}
  \begin{split}
    g_1 &= \sinh(\sqrt{2}y_o) [\cos(\alpha/2)-\sin(\alpha/2)] \\
      &- \sin(\sqrt{2}y_o)[\cos(\alpha/2)+\sin(\alpha/2)] ,
  \end{split} \\
  \begin{split}
    g_2 &= \sinh(\sqrt{2}(\pi-y_o)) [\cos(\alpha/2)+\sin(\alpha/2)] \\
      &+ \sin(\alpha/2-\sqrt{2}(\pi-y_o)) .
  \end{split}
\end{align}\end{subequations}
It is straightforward to check by inspection, that both $f_1$ and
$f_3$ are strictly positive and isolated from $0$.

The derivative of $X$ over $\alpha$ takes the form
\begin{equation}
  X'(y_o,\alpha) = \frac{f_1(y_o)f_3(y_o)}{2f_2(y_o)g_2^2(y_o,\alpha)} ,
\end{equation}
where the function
\begin{equation}\begin{split}
  f_3 &= \cosh(\sqrt{2}(\pi-2y_o)) - \cosh(\sqrt{2}\pi) \\
  &+ \cos(\sqrt{2}(\pi-y_o))\sin(\sqrt{2}y_o)-\sinh(\sqrt{2}y_o) \\
  &+ \sin(\sqrt{2}(\pi-y_o))\sin(\sqrt{2}y_o)+\sinh(\sqrt{2}y_o)
\end{split}\end{equation}
has (also by inspection) the following properties
\begin{itemize}
  \item $f_3(y_o=0) = 0$,
  \item $\forall y_o \in [10^{-3},\pi]\ f_3(y_o) \in [-50,-0.1]$,
  \item $\forall y_o \in [0,10^{-3}]\ [\partial_{y_o}f_3](y_o) \in [-123.0,-122.5]$,
\end{itemize}
which again implies its boundedness and isolation from zero for any
$0<y_o<\pi$.

The derivative $\partial_{\alpha}\beta$ can be expressed using
$X,X'$ as follows:
\begin{equation}\label{eq:app-diff}
  [\partial_{\alpha}\beta](\alpha,y_o) =
  \frac{X'(y_o,\alpha)}{1+X^2(y_o,\alpha)}
  = \frac{f_1 f_2 f_3}{f_1^2g_1^2+f_2^2g_2^2}
\end{equation}
and by the properties of $f_1,f_2,f_3$ it never vanishes and can
reach infinity only when both $g_1$ and $g_2$ vanish simultaneously.
The condition $g_1=g_2=0$ can be reexpressed as the linear system of
equations for $\sin(\alpha/2)$ and $\cos(\alpha/2)$ (treated as
independent variables). The existence of the nontrivial solution to
that system requires that the determinant of the equation matrix
vanishes, that is
\begin{equation}\begin{split}
  [\sinh(\sqrt{2}(\pi-y_o))+\cos(\sqrt{2}(\pi-y_o))]
    &\times [\sinh(\sqrt{2}y_o)-\sin(\sqrt{2}y_o)] \\
    + [\sinh(\sqrt{2}y_o)+\sin(\sqrt{2}y_o)]
    &\times [\sinh(\sqrt{2}(\pi-y_o))-\sin(\sqrt{2}(\pi-y_o))] =0 .
\end{split}\end{equation}
Since all the terms in the square brackets are explicitly positive
within $y_o\in]0,\pi[$ this condition is never satisfied. Therefore,
the denominator of the rightmost expression in \eqref{eq:app-diff}
never vanishes, so for fixed value of $y_o$ the derivative
$\partial_{\alpha}\beta$ is bounded due to the continuity of $X,X'$
and the compactness of the domain of $\alpha$. Thus we have
\begin{equation}
  0 > c \geq [\partial_{\alpha}\beta](y_o,\alpha) \geq C > -\infty .
\end{equation}

\section{Classical effective dynamics}\label{app:eff}

In Sec.~\ref{sec:lqc-dyn} we found that wave functions which start
out being sharply peaked on a classical trajectory in the low
curvature region remain sharply peaked throughout the evolution,
including the Planck regime. This strongly suggests that there may
be effective dynamical trajectories on the classical phase space
which incorporate the appropriate quantum corrections in the Planck
regime and approximate the full quantum evolution quite well
throughout evolution. A particularly convenient heuristic method to
arrive at these effective equations was proposed in \cite{sv} and
later derived more systematically analytically in the case
$\Lambda=0$ \cite{vt}. It was successfully tested in several cases
of isotropic LQC \cite{aps3,bp,acs,apsv}. Here we briefly recall how
these effective equations arise by adapting the detailed discussion
for the $\Lambda <0$ model from \cite{bp} to the $\Lambda
>0$ case now under consideration.

The strategy comes from a geometrical formulation of quantum
mechanics in which the quantum Hilbert space is regarded as an
infinite dimensional symplectic manifold, sometimes called `the
quantum phase space'. The idea is to find an embedding of the finite
dimensional classical phase space into this infinite dimensional
`quantum phase space' such that the full quantum evolution preserves
the image of the embedding to a good approximation. For a harmonic
oscillator, the embedding is provided by coherent states (whose
dispersions are determined by the mass and the spring constant).
Such an embedding is possible also for FLRW models \cite{vt,jw}(for
a brief summary, see \cite{as}). The result is a quantum corrected,
effective constraint. Although it is simply a function on the
classical phase space ---obtained by taking the expectation values
of the quantum constraint operator in states corresponding to the
image of the embedding--- it differs from the classical constraint
function by terms involving $\hbar$. As mentioned above, dynamical
trajectories generated by this effective Hamiltonian constraint have
turned out to provide an excellent approximation to the full quantum
evolution of states that start out to be sharply peaked around a
classical trajectory in the low curvature regime.

In the LQC literature, the effective Hamiltonian constraint is
written using lapse $N=1$ so that it generates evolution in proper
or cosmic time. Therefore, to facilitate comparison, we will do the
same here. Then, the effective Hamiltonian constraint is given by

\begin{equation}\label{eq:Heff}
  \Ham_{\eff} =
  -\frac{3}{8\pi G\gamma^2\bar{\mu}^2}\,|p|^{\frac{1}{2}}\,
  \sin^2(\bar{\mu}c)
  \,+\, \frac{1}{2}\frac{p^{2}_{(\phi)}}{|p|^{\frac{3}{2}}}
  \,+\, \frac{p^{\frac{3}{2}}}{8\pi G}\, \Lambda \approx 0\,  .
\end{equation}
By calculating the Hamilton's equations from $\Ham_{\eff}$, and
using \eqref{eq:Heff} again to simplify the resulting expressions,
one arrives to the following evolution equation for the energy
density $\rho$ of the scalar field:
\begin{equation}\label{eq:Fried1}
  \rho' = \pm\,4\sqrt{3\pi G} \left[ \rho\,\rho_{\rm tot}
  \left(1-\frac{\rho_{\rm tot}}{\rcr}\right)\right]^{1/2} .
\end{equation}
where $\rho_{\rm tot} := \rho+\Lambda/(8\pi G)$ and the prime
denotes the derivative with respect to the scalar field $\phi$.
(Thus, one first calculates the derivatives of $p$ and $\phi$
with respect to proper time by taking Poisson brackets of $p$
and $\phi$ with $\Ham_{\eff}$ and then combines them to find
$p'$ and then $\rho'$.) However this equation is inconvenient
to use in numerical simulations because it is not regular at
$\rho=0$ and $\rho_{\tot}=\rcr$. Also, the sign in front of the
righthand side changes in the process of evolution (at the
bounce where $\rho_{\tot}=\rcr$ and at the recollapse where
$\rho=0$). Therefore for the purpose of finding the solution it
is more convenient to use the second order equation derived
from \eqref{eq:Fried1},
\begin{equation}\label{eq:Fried2}
  \rho'' = 24\pi G \left[ (2\rho+\rho_{\tot}) \left( 1 -
  \frac{\rho_{\tot}}{\rho_c} \right)
   - \rho \right]\, ,
\end{equation}
which admits a unique global solution to the initial value problem
with the initial data $(\rho(\phi_o),\rho'(\phi_o))$,
where $\rho'(\phi_o)$ is determined from $\rho_{\tot}$ via
\eqref{eq:Fried1}.
\begin{figure}[hbt!]
  \includegraphics[width=3.2in,height=2.4in]{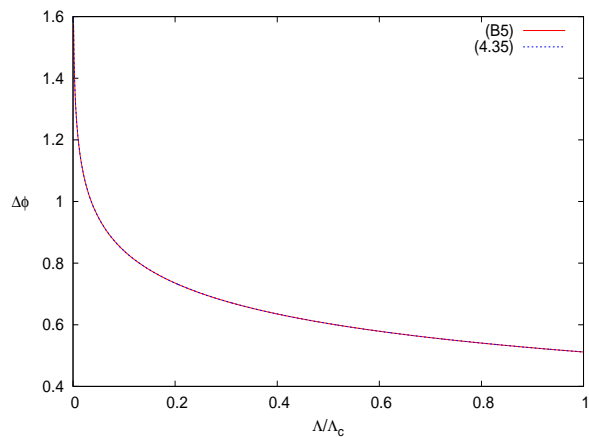}
  \caption{The period $\Delta\phi$ of the evolution predicted by
  the effective theory via \eqref{eq:eff-period} is compared against
  the approximate period of the genuine quantum evolution
  \eqref{eq:lqc-period}.}
  \label{fig:Dphi}
\end{figure}

Although this equation can be solved analytically, its solution
is expressed in terms of the elliptic integrals of the first
and second kind whose values have to be found numerically.
Therefore, it is more convenient to solve it numerically from
the beginning. In the actual calculations (such as the ones
presented in Fig.~\ref{fig:lqc-dynamics}) this step was carried
out with the use of the adaptive Runge-Kutta method of the
$5$th order known as \emph{RK45} (Cash-Carp).

The form of equation \eqref{eq:Fried2} implies already that the
trajectory $\rho(\phi)$ is periodic, with the period given by
a direct integration of \eqref{eq:Fried1}:
\begin{equation}\label{eq:eff-period}
  \Delta\phi = \frac{1}{\sqrt{12\pi G}}\,
  \int_{\rho=0}^{\rho_{\tot}=\rho_c} \rd \rho
  \left[ \rho_{\tot}\,\rho
  \left(1-\frac{\rho_{\tot}}{\rho_c}\right)\right]^{-1/2} .
\end{equation}
The dependence of $\Delta\phi$ on the value of cosmological constant
is presented in Fig.~\ref{fig:Dphi}, where its values were
calculated numerically via the standard trapezoid method. The
extremal points correspond respectively to the values $\rho=0$
(minimum corresponding to the recollapse at infinite volume) and
$\rho=\rcr-\Lambda/(8\pi G)$ (maximum corresponding to the quantum
bounce).


\begin{thebibliography}{99}

\bibitem{aps3} A.~Ashtekar, T.~Paw{\l}owski and P.~Singh, {Quantum
    nature of the big bang: Improved dynamics}, Phys. Rev. D {\bf
    74} 084003 (2006).

\bibitem{aps1} A.~Ashtekar, T.~Paw{\l}owski and P.~Singh, {Quantum
    nature of the big bang}, Phys. Rev. Lett. \textbf{96} 141301
    (2006). 

\bibitem{aps2} A.~Ashtekar, T.~Paw{\l}owski and P.~Singh, {Quantum
    nature of the big bang: An analytical and numerical
    investigation I}, Phys. Rev. D \textbf{73} 124038 (2006).

\bibitem{as}A.~Ashtekar and P.~Singh, Loop quantum cosmology:
    A status report, Class. Quant. Grav. \textbf{28} 213001 (2011).

\bibitem{warsaw-full} M.~Domaga{\l}a, K.~Giesel, W.~Kami\'nski and
    J. Lewandowski, Gravity quantized, Phys. Rev. D\textbf{82}
    104038 (2010)

\bibitem{consistent2}  D. Craig, P. Singh, Consistent
    probabilities in Wheeler-DeWitt quantum cosmology, Phys. Rev.
    D {\bf 82} 123526 (2010).

\bibitem{consistent3} D. Craig, P. Singh, Consistent histories
    in loop quantum cosmology, In preparation (2011).

\bibitem{bp} E.~Bentivegna and T.~Paw{\l}owski, Anti-deSitter
    universe dynamics in LQC,  Phys. Rev. D \textbf{77} 124025
    (2008).

\bibitem{acs}A.~Ashtekar, A.~Corichi and P.~Singh, {Robustness
    of predictions of loop quantum cosmology}, Phys. Rev.
    D \textbf{77} 024046 (2008).



\bibitem{awe1}A.~Ashtekar and E.~Wilson-Ewing, The covariant
    entropy bound and loop quantum cosmology, Phys. Rev. D
    \textbf{78} 06407 (2008). 

\bibitem{abl} A.~Ashtekar, M.~Bojowald and J.~Lewandowski,
    {Mathematical structure of loop quantum cosmology}. Adv. Theo.
    Math. Phys. \textbf{7} 233--268 (2003)

\bibitem{dm} D.~Marolf, Refined algebraic quantization:
    Systems with a single constraint. \texttt{arXiv: gr-qc/9508015}; \\
    {Quantum observables and recollapsing dynamics}. Class. Quant.
    Grav. {\bf 12} 1199--1220 (1994).

\bibitem{almmt} A.~Ashtekar, J.~Lewandowski, D.~Marolf,
    J.~Mour\~ao and T.~Thiemann, Quantization of diffeomorphism invariant
    theories of connections with local degrees of freedom. {Jour.
    Math. Phys.} \textbf{36} 6456--6493 (1995).

\bibitem{abc}A.~Ashtekar, L.~Bombelli and A.~Corichi, Semiclassical
    states for constrained systems, Phys. Rev. D \textbf{72} 025008
    (2005).

\bibitem{klp1}W.~Kami\'nski, J.~Lewandowski and T.~Paw{\l}owski,
    Physical time and other conceptual issues of QG on the example
    of LQC, Class. Quant. Grav. \textbf{26} 035012 (2009).

\bibitem{AbramowitzStegun} M.~Abramowitz and I.~Stegun,
    Handbook of Mathematical Functions With Formulas, Graphs and
    Mathematical Tables (U.S. Government Printing Office, Washington,
    1972).

\bibitem{warsaw3} W.~Kami\'nski and J.~Lewandowski, The flat FRW model
    in LQC: the self-adjointness, Class. Quant. Grav. \textbf{25} 035001
    (2008)

\bibitem{kp-posL} W.~Kami\'nski and T.~Paw{\l}owski, The LQC evolution
    operator of FRW universe with positive cosmological constant
    Phys. Rev. D \textbf{81} 024014 (2010).

\bibitem{ReedSimon-v2} M.~Reed and B.~Simon, Methods of Modern
    Mathematical Physics Vol. II (Academic Press, New York,
    1978).

\bibitem{kp-scatter} W.~Kami\'nski and T.~Paw{\l}owski, Cosmic recall
    and the scattering picture of loop quantum cosmology, Phys. Rev.
    D \textbf{81} 084027 (2010).

\bibitem{k-poLwdw} A.~Kreienbuehl, Singularity avoidance and
    time in quantum gravity, Phys. Rev. D \textbf{79} 123509 (2009).

\bibitem{mop} G.A.~Mena-Marug\'an, J.~Olmedo and T.~Paw{\l}owski,
    {Prescriptions in Loop Quantum Cosmology: A comparative analysis},
    Phys. Rev. D \textbf{84} 064012 (2011).

\bibitem{ar}A. Ashtekar and J.~D.~Romano, Spatial infinity as a
    boundary of space-time, Class. Quant. Grav. \textbf{9}, 1121-1150
    (1992).

\bibitem{rp} R.~Penrose, \textit{Cycles of Time: An Extraordinary
    New View of the Universe}  (Bodley Head, London, 2010).

\bibitem{hp-qg} V.~Husain and T.~Paw{\l}owski,
    Time and a physical hamiltonian for quantum gravity (2011),
    \texttt{arXiv:1108.1145};\\
    V.~Husain and T.~Paw{\l}owski,
    Loop quantum gravity with dust (in prep.).

\bibitem{hp-qc} V.~Husain and T.~Paw{\l}owski
    Dust reference frame in quantum cosmology,
    Class. Quant. Grav. \textbf{28}, 225014 (2011);\\
    V.~Husain and T.~Paw{\l}owski,
    Loop Quantum Cosmology in dust frame (in prep.).

\bibitem{klp2} W.~Kami\'nski, J.~Lewandowski and T.~Paw{\l}owski,
    Quantum constraints, Dirac observables and evolution: group
    averaging versus Schroedinger picture in LQC, Class. Quant. Grav.
    \textbf{26} 245016 (2009).

\bibitem{k-pres} W.~Kami\'nski, Schroedinger Picture vs Group Averaging in LQC,
    Loops'11 conference presentation (2011);\\
    W.~Kami\'nski, Matematyczne w{\l}asno\'sci wi\c{e}z\'ow kwantowych
      (Mathematical properties of quantum constraints). Ph.D.
    Dissertation, Warsaw University (2011).

\bibitem{sv} P.~Singh and K.~Vandersloot, Semi-classical states,
    effective dynamics and classical emergence in loop quantum
    cosmology, Phys. Rev. D \textbf{72} 084004 (2005).

\bibitem{vt} V.~Taveras, LQC corrections to the Friedmann equations
    for a universe with a free scalar field, Phys. Rev. D
    \textbf{78} 064072 (2008). 

\bibitem{apsv} A.~Ashtekar, T.~Paw{\l}owski, P.~Singh and
    K.~Vandersloot, {Loop quantum cosmology of k=1 FRW
    models}. Phys. Rev. D \textbf{75} 0240035 (2006).


\bibitem{jw} J.~Willis, {On the low energy ramifications and a
    mathematical extension of loop quantum gravity}. Ph.D.
    Dissertation, The Pennsylvania State University (2004).

\end{thebibliography}
\end{document}